\documentclass[fleqn,usenatbib]{mnras}
\usepackage[utf8]{inputenc}
\usepackage{amsfonts}
\usepackage{graphicx}
\usepackage{grffile}
\usepackage{amsmath}
\usepackage{amssymb}
\usepackage{newtxtext}%,newtxmath}
\usepackage[capitalise]{cleveref}
\usepackage{lmodern}
\usepackage{hyperref}
\usepackage{array}
\usepackage{listings}
\usepackage{comment}
\usepackage{bm}
\usepackage{slashed}
\usepackage{mathtools}

\usepackage{siunitx}
\usepackage{multirow}
\usepackage{booktabs}
\usepackage[para]{threeparttable}
\usepackage{ulem}
\usepackage{xcolor}
\newcommand{\be}{\begin{equation}}
\newcommand{\ee}{\end{equation}}
\newcommand{\bea}{\begin{eqnarray}}
\newcommand{\eea}{\end{eqnarray}}
 
\def\ba#1\ea{\begin{align}#1\end{align}}

\renewcommand{\P}{\mathcal{P}}

\def\tt{\tilde t_{0}}

\definecolor{RedWine}{rgb}{0.743,0,0}
\definecolor{GrassGreen}{rgb}{0.125,0.75,0.125}
\definecolor{RoyalBlue}{rgb}{0.25,0.41,0.88}
\definecolor{DarkCyan}{rgb}{0,0.5,0.5}

\newcommand{\cc}{{\cal C}}

\newcommand{\hto}{{\rm H_2}}

\newcommand{\hd}{{\rm HD}}
\newcommand{\rH}{{\rm H}}

\newcommand{\re}{{\rm e}}
\numberwithin{equation}{section}

\usepackage{tikz}
\usetikzlibrary{shapes.geometric, arrows}
\tikzstyle{startstop} = [rectangle, rounded corners, minimum width=1cm, minimum height=0.5cm,text centered, draw=black, fill=red!30]
\tikzstyle{io} = [rectangle, minimum width=1cm, minimum height=0.5cm, text centered, draw=black, fill=blue!30]
\tikzstyle{process} = [rectangle, minimum width=1cm, minimum height=0.5cm, text centered, draw=black, fill=orange!30]
\tikzstyle{decision} = [rectangle, minimum width=1cm, minimum height=0.5cm, text centered, draw=black, fill=green!30]
\tikzstyle{arrow} = [thick,->,>=stealth]
\tikzstyle{narrow} = [thick,-,>=stealth]

\title[Bridging the Pop.~III Scales ]{Towards A Universal Analytical Model of Population III Star Formation: A Bridge Between Cosmological Scales and Protostars}

\author[J. Gurian et al.]{James Gurian\textsuperscript{\href{https://orcid.org/0000-0002-8677-1038}{\includegraphics[width=2.5mm]{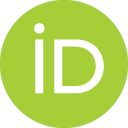}}\,}$^1$,\thanks{E-mail: jgurian@perimeterinstitute.ca}
Boyuan Liu\textsuperscript{\href{https://orcid.org/0000-0002-4966-7450}{\includegraphics[width=2.5mm]{orcid.png}}\,}$^{2}$,
Donghui Jeong\textsuperscript{\href{https://orcid.org/0000-0002-8434-979X}{\includegraphics[width=2.5mm]{orcid.png}}\,}$^{3,4}$,
Takashi Hosokawa\textsuperscript{\href{https://orcid.org/0000-0003-3127-5982}{\includegraphics[width=2.5mm]{orcid.png}}\,}$^5$,
Shingo Hirano\textsuperscript{\href{https://orcid.org/0000-0002-4317-767X}{\includegraphics[width=2.5mm]{orcid.png}}\,}$^{6,7}$,
\newauthor 
Volker Bromm\textsuperscript{\href{https://orcid.org/0000-0003-0212-2979}{\includegraphics[width=2.5mm]{orcid.png}}\,}$^{8,9,10}$, 
 and Naoki Yoshida\textsuperscript{\href{https://orcid.org/0000-0001-7925-238X}{\includegraphics[width=2.5mm]{orcid.png}}\,}$^{11,12,13}$\\
$^1$ Perimeter Institute for Theoretical Physics, Waterloo, Ontario, N2L 2Y5, Canada\\
$^2$ Institut für Theoretische Astrophysik, Zentrum für Astronomie, Universität Heidelberg, Albert Ueberle Straße 2, D-69120 Heidelberg, Germany\\
$^3$ Department of Astronomy and Astrophysics and Institute for Gravitation and the Cosmos,\\ ~The Pennsylvania State University, University Park, PA, 16802, USA\\
$^4$School of Physics, Korea Institute for Advanced Study (KIAS), 85 Hoegiro, Dongdaemun-gu, Seoul, 02455, Republic of Korea\\
$^5$Department of Physics, Kyoto University, Sakyo, Kyoto 606-8502, Japan\\
$^6$Department of Applied Physics, Faculty of Engineering, Kanagawa University, Kanagawa 221-0802\\
$^7$Department of Astronomy, School of Science, University of Tokyo,  7-3-1 Hongo, Bunkyo, Tokyo 113-0033, Japan\\
$^{8}$Department of Astronomy, University of Texas, Austin, TX 78712, USA\\
$^{9}$Weinberg Institute for Theoretical Physics, University of Texas, Austin, TX 78712, USA\\
$^{10}$Cosmic Frontier Center, The University of Texas at Austin, Austin, TX 78712, USA\\
$^{11}$Department of Physics, School of Science, The University of Tokyo, 7-3-1 Hongo, Bunkyo, Tokyo 113-0033, Japan\\
$^{12}$Research Center for the Early Universe, School of Science, The University of Tokyo, 7-3-1 Hongo, Bunkyo, Tokyo 113-0033, Japan\\
$^{13}$Kavli Institute for the Physics and Mathematics of the Universe (WPI), The University of Tokyo, Kashiwa, Chiba 277-8583, Japan\\
}
\date{Accepted XXX. Received YYY; in original form ZZZ}
\pubyear{2026}
\begin{document}

\label{firstpage}
\pagerange{\pageref{firstpage}--\pageref{lastpage}}
\maketitle

\begin{abstract}
We construct an analytical model of Population III star formation that connects the cosmological radiation background to sub-AU protostellar disk fragmentation, a dynamic range inaccessible to any single simulation. Our approach is based on combining separate models of the disparate relevant scales: from the cosmological environment to the host-halo scale, from the halo scale to the scale of the star-forming cloud, and from the cloud scale to the fragmenting, accreting protostellar disk. Individually and collectively, the models agree well with the predictions of state of the art simulations, while remaining computationally inexpensive and physically transparent. As an example of the applicability of the model, we study the effects of varying the Lyman-Werner flux on the Pop.~III star formation efficiency. We show that depending on the halo properties and the strength of the dissociating radiation field, the halo-scale Pop.~III star formation efficiency varies by more than two orders of magnitude from $\varepsilon_{\rm SFE,H} \approx 10^{-3}$ to  $\varepsilon_{\rm SFE, H} \approx 0.5$. The abrupt transitions between hydrogen-deuteride cooling (in low virial temperature mini-halos subjected to low radiation backgrounds), molecular hydrogen cooling (at intermediate temperatures and radiation intensities), and atomic cooling (in higher temperature halos exposed to strong radiation fields) produces sharp features in the halo-scale star formation efficiency as a function of the halo properties. Meanwhile, at the scale of individual star-forming clouds, the star formation efficiency is $\varepsilon_{\rm SFE,c} \gtrsim 0.2$. That is, pristine gas in a halo is converted into unstable clouds at a wide range of efficiencies, and these unstable clouds are efficiently converted into Pop.~III stars. 
\end{abstract}


\section{Introduction}
The first generation, Population~III (Pop.~III) stars form from pristine, metal-free gas \citep[e.g.,][]{Haiman1996,Tegmark1997}. Their formation is a critical phase in the evolution of the Universe, marking the end of the Cosmic Dark Ages, and providing the metals which catalyze subsequent generations of star and galaxy formation \citep[e.g.,][]{Karlsson2013}. Due to their rarity and high formation redshift, imaging of a single Pop.~III star is challenging in the near term \citep[e.g.,][]{Zackrisson24,Schauer2020,Schauer2022}. However, the James Webb Space Telescope (JWST) could potentially detect a Pop.~III cluster \citep{Trussler23,Cai2025,Fujimoto25,Fujimoto2025,Morishita:2025zvd}, and has already detected intriguing, extremely metal-poor galaxies \citep{Vanzella_2023,Cullen25,hsiao2025,Maiolino_Hebe2026}.  The properties of these stars could also leave an imprint on the 21~cm temperature and power spectrum \citep[e.g.,][]{Gessey_Jones_2022,Cruz_2025,Liu2025}. More locally, metal-poor stars in the Milky Way neighborhood, through their abundance patterns, provide important clues about the properties of the first-generation stars \citep{Beers_2005,Koutsouridou24,Rossi2025,Chiti2025,Frebel2026}.

An intensive simulation campaign focused on the small-scale physics of Pop.~III star formation has largely converged on a ``standard model'' of Pop.~III star formation \citep[e.g.,][]{Bromm:2013,Klessen2023stars}. In this model, the gas in high-redshift $z\approx 10$--$20$, low mass $\sim 10^5$--$10^8 \, \rm M_\odot$ halos condenses (typically) monolithically,\footnote{Studies find that in regions of high baryon-dark matter streaming velocity, the gas can instead be subject to turbulent fragmentation \citep{Hirano_2018,Hirano2025}.} resulting in the formation of a gravitationally unstable cloud, whose mass is dictated primarily by the dominant coolant. These clouds range in mass from as low as a few tens of solar masses (in hydrogen-deuteride, $\hd$ cooled clouds) through $\sim 1000 \, \rm M_\odot$ (in molecular hydrogen, $\hto$ cooled clouds) all the way to  $\gtrsim10^5 \, \rm M_\odot$ in neutral, atomic hydrogen cooled halos. The cloud then collapses to form a proto-stellar accretion disk, which fragments into a growing cluster. The cluster grows until the accretion is ultimately terminated by (proto)-stellar feedback. 

Simulations almost universally find that the mass spectrum produced by fragmentation in the proto-stellar disk is top heavy, with most of the stellar mass concentrated near the high end of the distribution \citep{stacy_constraining_2013,Wollenberg_2020,Prole2022,Klessen2023stars, Sharda_2025}. Key open questions include the lower limit of the Pop.~III mass function \citep{Ishiyama2016,hirano_formation_2017,Becerra2018,nishijima2023lowmasspopiiistar}, the importance of magnetic fields in the evolution \citep{Saad_2022, Sharda2025,Machida2025}, and the mechanism by which ionizing photons first break out of the dense gas surrounding the protostars \citep[e.g.][]{Hosokawa_2016,Jaura_2022,Nebrin2025}. However, the following predictions appear to be quite robust: 1) the mass function is top heavy, 2) a significant fraction of the mass present in each gravitationally unstable cloud is converted into stars, and 3) the range of cloud masses is very wide. 

This picture makes clear that Pop.~III star formation entails enormous dynamic range, from the $>\rm Mpc$ large-scale environment which dictates the chemical-thermal properties of the collapsing gas to the fragmentation cascades which terminate on $< \rm AU$ scales. Simultaneously resolving all of the relevant physics in a simulation is intractable in the foreseeable future. In present day star formation, this difficulty can be sidestepped due to the high degree of regularity in the stellar initial mass function (IMF) \citep{Hennebelle24}. Although the IMF is not exactly universal, it depends relatively weakly on the large-scale environment. Thus, adopting a universal IMF in simulations and semi-analytic treatments of galaxy formation is often a reasonable approximation. In contrast, for Pop.~III stars, given the enormous range in cloud masses (determined by the cosmological initial conditions) and the top heavy mass function, Pop.~III star formation is likely to be highly sensitive to the large-scale environment. That is, the global (cosmic-averaged) Pop.~III IMF is unlikely to resemble the local IMF in any given Pop.~III cluster due to the importance of the environment dependent high-mass cutoff. 

From an observational perspective, this is a crucial point. On one hand, accurate interpretation of the many observational probes of Pop.~III stars which involve surveying cosmological volumes requires taking into account the complex environmental dependencies of Pop.~III star formation. On the other, methods such as stellar archaeology which sample only the Milky Way neighborhood cannot directly constrain the cosmic averaged Pop.~III properties.

Thus, computationally inexpensive tools which capture the wide range of Pop.~III star formation outcomes are a necessary ingredient to connect the lessons learned from cutting edge small-scale simulations to large-volume theoretical studies and observational searches. In pursuit of this goal, we have developed analytic models of Pop.~III star formation at the halo scale \citep{gurian2024zero}, cloud scale \citep{Gurian2025}, and disk scale \citep{liu2024}. Here, we integrate these models into an end-to-end semi-analytic pipeline for modeling Pop.~III star formation. This combined model captures the influence of the large-scale environment on the outcomes of Pop.~III star formation. Further, via a small number of physical parameters this model facilitates marginalization over the remaining uncertainties in the small-scale processes. 

This paper is organized as follows. In \cref{sec:model_spec} we develop successively the halo (\cref{sec:haloscale}), cloud (\cref{sec:cloud_scale}), and disk (\cref{sec:cluster_evo}) models. In \cref{sec:final_model} we calculate simple estimates of which halos may form Pop.~III stars and present the results of the integrated model. In \cref{sec:sfe} we apply the model to a wider range of parameters, and study the dependence of the Pop.~III star-formation efficiency on the halo properties and the Lyman-Werner radiation field. We conclude with \cref{sec:discussion}. 

\section{Model Specification and Validation}
\label{sec:model_spec}
\begin{figure}
    \centering
    \scriptsize
\begin{tikzpicture}[node distance=1.5cm]
%\centering
\node (cosmic) [startstop, align=left] {\textbf{Cosmological context ($\rm \mathcal{O}(10)-\mathcal{O}(10^4)\ cMpc$)}\\ 
Cosmological (hydrodynamic) simulation, halo merger tree/mass function};
\node (halo) [process, below of=cosmic,xshift=0cm, yshift=0cm, align=left] {\textbf{Halo scale ($\sim 0.1-1\ \rm kpc$)}\\ Input: halo mass ($M_H$), redshift ($z$), LW intensity ($J_{21}$) \\ 
Output: density-temperature ($n$-$T$) evolution track of collapsing gas};
\node (cloud) [decision, below of=halo, align=left] {\textbf{Cloud scale ($\sim 1-10\ \rm pc$)}\\ Input: $n$-$T$ track, turbulent Mach number ($\mathcal{M}$), spin parameter ($\lambda$) \\ Output: gas infall history onto the disk+protostar system ($\dot{M}_{\rm in}$)};
\node (disk) [io, below of=cloud, xshift=0cm, align=left] {\textbf{Disk scale ($\sim 1-10^5\ \rm AU$)} \\ Input: $\dot{M}_{\rm in}$, lower limit ($m_1$), (time-evolving) upper limit ($m_2$),\\ and power-law index ($\alpha$) of the stellar mass function\\ Output: total stellar mass (per cloud), IMF, star formation efficiency ($\epsilon_{\rm SFE}$)};
\node (cd)[below of=cosmic, xshift=3.4cm,yshift=0.82cm]{Star-forming clouds$\ \ \ \ $};
\node (fdbk)[below of=cloud, xshift=-3.7cm,yshift=0.8cm]{Stellar feedback};
\draw [arrow] (halo) -- (cloud); % node[xshift=1cm]{$T$-$n$ track} 
\draw [arrow] (cloud) -- (disk);
\draw [dashed,arrow] (cosmic) -- node[xshift=-0.9cm,yshift=0cm]{Star-forming halos} (halo);
\draw [dashed,narrow] (cosmic) -| (cd);
\draw [dashed,arrow] (cd) |- (cloud);
\draw [dashed,narrow] (disk) -| (fdbk);
\draw [dashed,arrow] (fdbk) |- (cloud);
\draw [dashed,arrow] (fdbk) |- (halo);
\draw [dashed,arrow] (fdbk) |- (cosmic);
\end{tikzpicture}
    \caption{Connections between the cosmological context (red) and the three models of relevant scales in this work: the halo scale (orange), the cloud scale (green), and the disk scale (blue). The solid arrows show the workflow of our combined model: The output of a larger-scale model serves as key input for the subsequent smaller-scale model, and a minimal set of additional free parameters are considered to capture diverse Pop~III star formation pathways. The dashed arrows demonstrate how our analytical models can be integrated into large-scale cosmological simulations and semi-analytical models based on halo merger trees or mass functions. }
    \label{fig:pop3_flow_chart}
\end{figure}
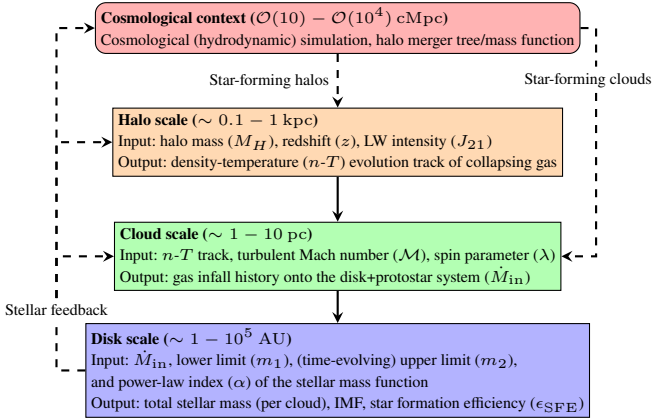

The model of Pop.~III star formation described in this work is composed of three separate components corresponding to the relevant scales in the problem: the halo, cloud, and disk. In this section, we introduce and link these models. The scales and their connections are illustrated in \cref{fig:pop3_flow_chart}. The halo-scale model uses a ``one-zone'' model to predict the relationship between density and temperature for the collapsing gas starting from the virial density and temperature, based on \cite{gurian2024zero}. One-zone models follow the chemical-thermal evolution of a uniform-density patch of gas whose density is changing on some specified timescale (in this work, always a multiple of the local dynamical timescale). The density-temperature relationship calculated in the one-zone model defines an effective equation of state for the collapsing gas which, together with the rotation parameter and turbulent Mach number (defined below), are the inputs to the cloud-scale model, based on \cite{Gurian2025}. In turn, the cloud-scale model estimates the spherically averaged density profile of the gas over the course of the pre-stellar collapse and what we term the ``cloud-scale infall rate'', which is the time-dependent rate at which mass is deposited near the center of the cloud after the formation of the first protostar. The cloud-scale infall rate is the rate at which the gravitationally unstable cloud feeds the rotationally supported protostellar accretion disk. Finally, the disk model based on \cite{liu2024} takes as inputs the cloud-scale infall rate and parameters describing the mass spectrum of the protostellar fragments and the efficiency of angular momentum transport and ionizing feedback. This model predicts the distribution of stellar masses at the end of the accretion phase. 

 All model free parameters and their fiducial values are listed in \cref{tab:fid_pars}, while the parameters held fixed and used to calibrate the model against simulations are listed in \cref{tab:calib_pars}.

\begin{table}
    \centering
    \begin{tabular}{ccccccc}
        \toprule
        \multicolumn{3}{c}{Halo Scale} & \multicolumn{2}{c}{Cloud Scale } & \multicolumn{2}{c}{Disk Scale} \\
        \cmidrule(lr){1-3} \cmidrule(lr){4-5} \cmidrule(lr){6-7}
        $z$ & $M_{  H}\, [\mathrm{M_\odot}]$ & $J_{21}$ & $\mathcal{M}$ & $\lambda$ & $m_1\ [\mathrm{M_\odot}]$ & $\alpha$ \\
        \midrule
        20 & $5 \times 10^{5}$ & 0.0 & 1.0 & 0.15 & 1 & -1 \\
        \bottomrule
    \end{tabular}
    \caption{Model free parameters and their fiducial values. Halo scale: redshift ($z$), halo mass ($M_H$), Lyman-Werner intensity ($J_{21}$). Cloud scale: turbulent Mach number ($\cal M$), spin parameter ($\lambda$). Disk scale: lower limit of mass function ($m_1$), mass function power law index ($\alpha$).}
    \label{tab:fid_pars}
\end{table}
\begin{table}
    \centering
    \begin{tabular}{ccccccc}
        \toprule
        \multicolumn{3}{c}{Cloud Scale Parameters} & \multicolumn{4}{c}{Disk Scale Parameters} \\
        \cmidrule(lr){1-3} \cmidrule(lr){4-7}
        $\beta$ & $n_0\, [\mathrm{cm}^{-3}]$ & $\xi$ & $\epsilon$ & $\gamma_{\rm eff}$ & $r_{d, \, \rm max}\, [\mathrm{au}]$  & $\chi$\\
        \midrule
        1.0 & $10^3$ & 4 & 0.5 & 1.09 & $10^5$ & 0.6 \\
        \bottomrule
    \end{tabular}
    \caption{Calibrated model parameters. Cloud scale: acceleration factor ($\beta$), rotation transition density ($n_0$), and centrifugal support factor ($\xi$). Disk scale: accretion efficiency ($\epsilon$), effective adiabatic index ($\gamma_{\rm eff}$), maximum disk radius ($r_{d, \, \rm max}$), and accretion fraction at the upper limit of the mass function ($\chi$).}
    \label{tab:calib_pars}
\end{table}

In the remainder of this section, we describe the three component models separately, beginning with the halo scale (Section~\ref{sec:haloscale}), proceeding to the cloud scale (Section~\ref{sec:cloud_scale}), and concluding with the disk scale (Section~\ref{sec:cluster_evo}). Finally, we present the results of the combined model which illustrate the cosmological environment dependence of Pop.~III star formation (\cref{sec:final_model}). 

\subsection{Halo Scale}
\label{sec:haloscale}
We describe here the dependence of the chemical-thermal evolution in a pristine, collapsing cloud on the large-scale environment. The density-temperature track determined by this evolution will be the essential input for the subsequent modeling of the cloud and disk scales. 

The chemical-thermal evolution is determined using a one-zone model. That is, we solve the system of ordinary  differential equations which specify the temperature and chemical composition of a uniform density parcel of gas evolving on a specified timescale. This chemical-thermal evolution is calculated using \texttt{KROME} \citep{grassi_krome_2014}. We use the reaction network \texttt{react\_primordial\_wD} packaged within \texttt{KROME}, with modifications described below.

\begin{table}
    \centering
    \begin{tabular}{c|l|l}
    \hline\hline
        $k_{\rm H,9}$ & $\rm H + e^{-} \rightarrow H^-$& \cite{deJong1972} \\
        $k_{\rm H,10}$ & $\rm H^- + H \rightarrow H_2 + e^{-}$ & \cite{Krecker2010}
        \\
        $k_{\rH, 25}$ & $\rm H^- + \gamma \rightarrow H + e^{-}$& \cite{Shang2010} \\
        $k_{\rm H, 27}$ & $\rm H_2 + \gamma \rightarrow H + H$&\cite{Shapiro1987}  \\
                $k_{\rm D, 27}$ & $\rm D_2 + \gamma \rightarrow D + D$&\cite{Shapiro1987}\\
    \hline\hline
    \end{tabular}
    \caption{The reactions used to calculate $f$ (first three rows) and the additions to the \texttt{react\_primordial\_wD} network to include a Lyman-Werner radiation field(last three rows). The $\hd$ dissociation rate is assumed the same as the $\hto$ dissociation rate. The $\hto$ and $\hd$ dissociation rates are multiplied by the shielding function of \protect\cite{Wolcott_Green_2011}.  }
    \label{tab:reactions}
\end{table}
The halo properties enter first through the initial conditions for the chemical-thermal network, because the halo mass and redshift set the virial density and temperature. We initialize the one-zone model at the virial density reduced by a factor of 3: $n_i = n_{\rm vir}/3$, and temperature reduced by a corresponding factor assuming adiabatic evolution $T_i = T_{\rm vir}/3^{2/3}$, which allows for chemical reactions before virialization. The initial chemical abundances are set according to their global freeze-out values as in \cite{gurian2024zero, Gurian2025}. The redshift also sets the CMB temperature floor, which is implemented in \texttt{KROME} by modifying the cooling rates $\Lambda$ to the effective rate $\Lambda_{\rm eff}(T) = \Lambda(T) -\Lambda(T_{\rm CMB})$.

We augment the one-zone modeling of \cite{gurian2024zero,Gurian2025} by including the strength of the dissociating Lyman-Werner radiation as a model parameter via $J_{21} \equiv J_{\rm LW}/ (10^{21} \, \rm erg \,s^{-1}\, cm^{-2} \, Hz^{-1} sr^{-1})$, with $J_{\rm LW}$ the radiation intensity in the Lyman-Werner band. A weak cosmological Lyman-Werner background $J_{21} \lesssim 1$ is expected to develop as early star formation proceeds towards reionization \citep{Visbal_2014}, while a stronger Lyman-Werner intensity typically requires a nearby bright source \citep{Omukai01}. In order to include the Lyman-Werner radiation in our modeling, we add the reactions listed in \cref{tab:reactions} and the shielding fitting function of \cite{Wolcott_Green_2011} to the chemical network. The column density for the self-shielding is calculated using the number density and the Jeans length implied by the one-zone model \citep{grassi_krome_2014}. We assume a blackbody spectrum at $10^4 \, \rm K$ for the Lyman-Werner radiation, commonly referred to as a $T_4$ spectrum, for all of the photo-chemistry reactions. 

Given these initial conditions, the density is evolved on a characteristic collapse timescale $\dot \rho = \rho /t_{\rm col}$, adopting $t_{\rm col} = (f t_{\rm ff})$, where $f$ is a delay parameter and $t_{ff}$ is the free-fall timescale. The delay parameter is necessary because in some cases, a lack of coolants or strong rotational support can slow down the collapse compared to free-fall \citep{hirano_one_2014,gurian2024zero}, which promotes the formation of appreciable $\hd$. The is in turn significantly reduces the temperature compared to $\hto$ cooling alone, because $\hd$ has a permanent dipole moment. 

In this work we do not consider the dependence of $f$ on rotation or Mach number (cloud-scale quantities) but assume that $f$ can be predicted at the halo scale. In \cite{gurian2024zero} we estimated $f$ as a function of host halo mass and star formation redshift based on the $\hto$ production timescale. We here make a closely related argument, connecting the $\hto$ production time to the canonical Rees-Ostriker condition. The Rees-Ostriker criterion is an estimate for the onset of the runaway collapse \citep{Rees77}:
\begin{equation}
    t_{\cal C}(T,n, \vec x) \leq t_{\rm ff},
    \label{eq:ro}
\end{equation}
where the cooling timescale $t_{\cal C}$ is the time to radiate the thermal energy and depends on the temperature ($T$), density $n$, and chemical composition $\vec x$. We estimate the time at which the gas in a halo becomes Rees-Ostriker unstable by assuming that before the instability the density and temperature remain approximately constant at their virial values $T=T_{\rm V}$, $n=n_{\rm V}$ while the chemical composition $\vec x$ evolves.

For the chemical composition, we estimate the time-dependent molecular fraction by noting that for the $T_4$ spectrum we consider, the impact of the Lyman-Werner radiation is dominantly by photodetachment of $\rm H^-$. Therefore, an estimate of the $\hto$ abundance is 
\begin{equation}
    x_{\hto}(t) = k_{\rH,9} \frac{k_{\rm H, 10}}{k_{\rm H,10} + k_{25}(J_{21})}x_e n_{\rm V} t,
    \label{eq:h2_lw}
\end{equation}
where $k_{\rm H,9}$ is the rate for $\rm H + e^- \rightarrow H^-$ (which is the rate limiting step in the dominant $\hto$ production pathway), $k_{10}$ is the rate of $\rH^- + \rH \rightarrow \hto + e^-$ and $k_{25}$ is the $\rH^-$ photo-detachment rate of \cite{Shang2010}. Additionally,  $x_e$ is the cosmological freeze-out free electron fraction, which we take to be constant (i.e.~we neglect electron depletion). Throughout, the freeze-out chemical abundances, virial density and temperature, and reaction rates are as in \cite{gurian2024zero}. With this time-dependent $\hto$ fraction, we can solve for the time at which \cref{eq:ro} is satisfied, $t_{\rm RO}$. Then, we take $f=t_{\rm RO}/t_{\rm ff}$, restricted to $1 < f< 10$. The lower limit is because the collapse cannot be accelerated much faster than the dynamical timescale, while the upper is due to the $\hd$ production saturating by $f\gtrsim 5$. We have checked that computing $f$ by solving the full chemical network at constant (virial) density and temperature until $t_{\cc} < t_{\rm ff}$ produces similar results.

\begin{figure}
    \centering
    \includegraphics[width=\linewidth]{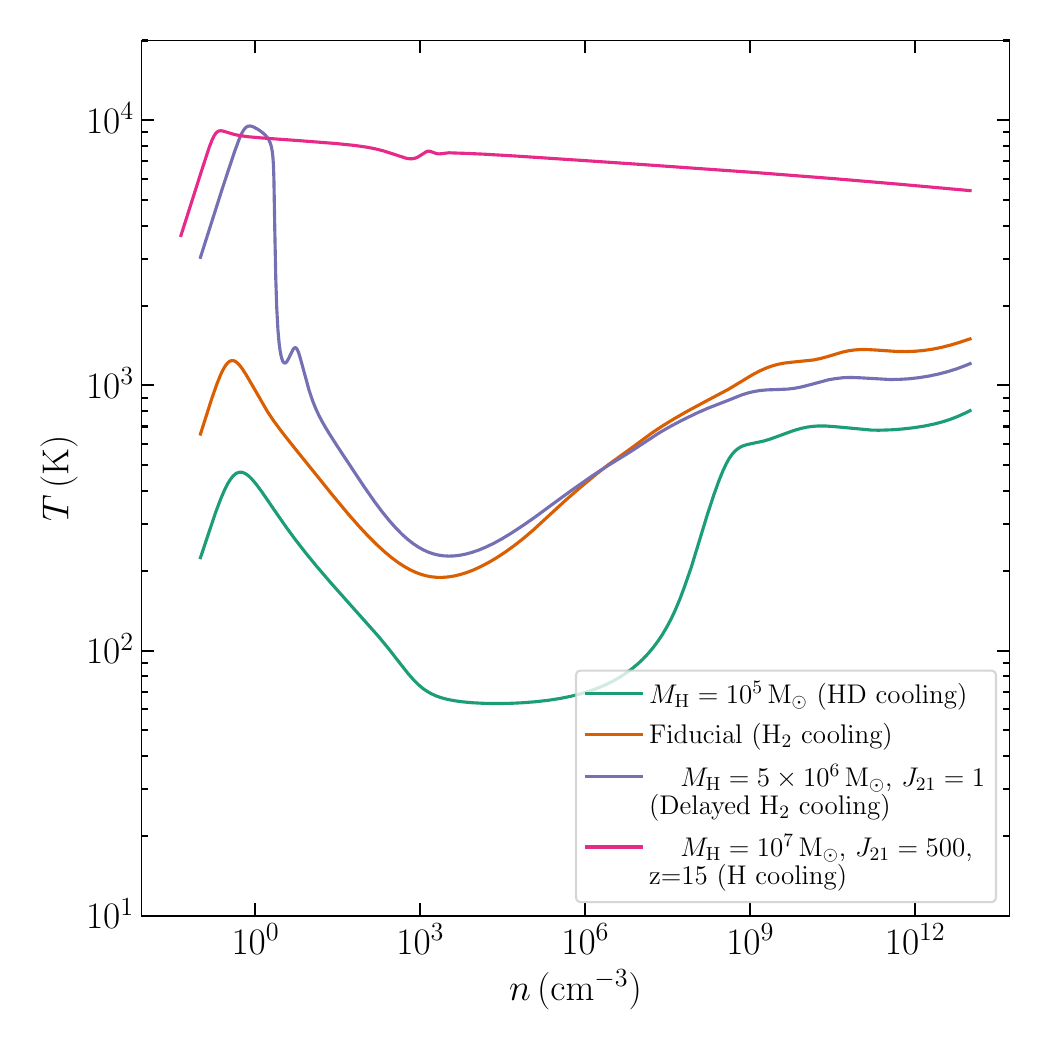}
    \caption{The density-temperature evolution for four cases which illustrate the possible cooling channels, labeled by the parameters which differ from their fiducial values (\cref{tab:fid_pars}). From lowest temperature minimum to highest: a $10^5 \, \rm M_\odot$ halo at $z=20$ with $J_{21}=0$ ($\rm HD$ cooling), a $5\times 10^5 \, \rm M_\odot$ halo at $z=20$ with $J_{21}=0$ (canonical $\hto$ cooling), a $5\times 10^6 \, \rm M_\odot$ halo at $z=20$ with $J_{21}=1$ (delayed $\hto$ cooling), and a $10^7 \, \rm M_\odot$ halo at $z=15$ with $J_{21}=500$ ($\rH$ cooling).  }
    \label{fig:n-t_tracks}
\end{figure}

In \cref{fig:n-t_tracks} we show the thermal evolution for four halos which illustrate the different typical cooling modes. Previous work shows that large $f$ activates the $\hd$ cooling mode which leads to low cloud masses, while large $J_{21}$ leads to atomic cooling as the $\hto$ is dissociated. \footnote{The exact value of $J_{21}$ necessary to shut down $\hto$ cooling has been the subject of extensive research \citep[e.g.,][]{Woods2019}. For the $T_4$ spectrum we consider, most studies place this critical value $30 \lesssim J_{\rm crit} \lesssim 300$ \citep{Sugimura2014, Latif14, Shang2010}. Adopting the same initial conditions as \cite{Latif14,Shang2010}, we find $J_{\rm crit} \approx 180$, compared to 30--40 for similar one-zone models in those works. The discrepancy can be at least partially traced to our adoption of the updated rate for the reaction $\rH^- + \rH \rightarrow \hto + \re^-$ from \cite{Krecker2010}, which is a factor of few larger than the rate adopted by those works at the relevant temperatures.} At intermediate $J_{21}$, the $\hto$ or $\hd$ channels can operate, but with a typically large delay parameter due to the impeded $\hto$ formation. Thus, we choose a $10^5 \, \rm M_\odot$ halo at $z=20$ with $J_{21}=0$ ($\rm HD$ cooling), a $5\times 10^5 \, \rm M_\odot$ halo at $z=20$ with $J_{21}=0$ (canonical $\hto$ cooling), a $5\times 10^6 \, \rm M_\odot$ halo at $z=20$ with $J_{21}=1$ (delayed $\hto$ cooling), and a $10^7 \, \rm M_\odot$ halo at $z=15$ with $J_{21}=500$ ($\rH$ cooling). For the canonical $\hto$ cooling case, the calculated value of the delay parameter is $f=1$, while for all other cases $f=10$.  

\begin{figure*}
    \centering
    \includegraphics[width=\linewidth,trim={0 6cm 0 0},clip]{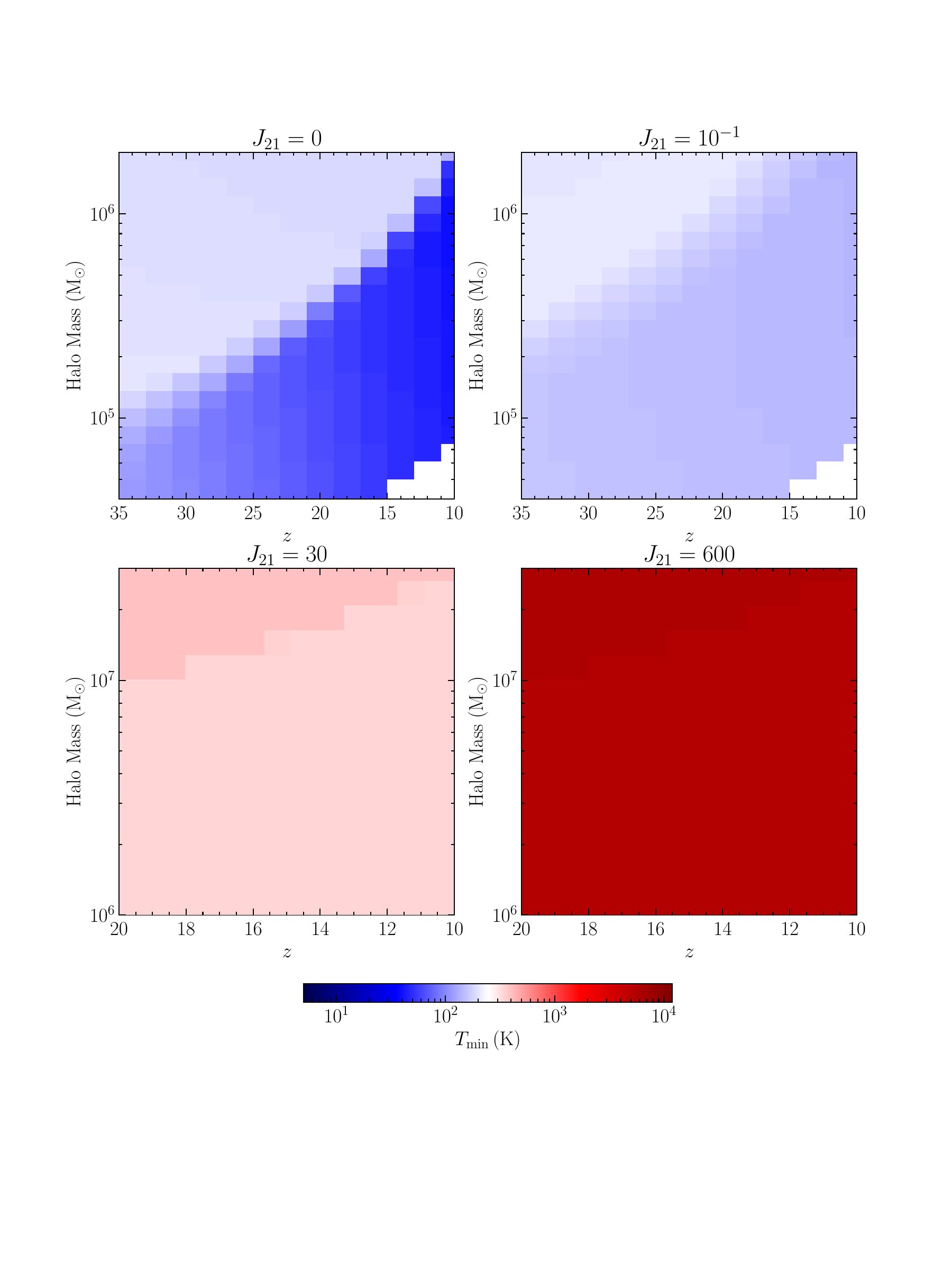}
    \caption{The minimum temperature of the collapsing gas for various halo masses, redshifts, and radiation field intensities. Note that white corresponds approximately to the minimum temperature attainable by $\hto$ cooling, so that dark blue indicates $\hd$ cooling while light red indicates  weak molecular cooling and dark red, atomic cooling. In the bottom right corner of the top panels, the virial temperature is so low that $\hto$ transitions cannot cool the halo and $t_{\rm RO}$ cannot be calculated. }
    \label{fig:T_min}
\end{figure*}

The temperature minimum in the collapsing gas has long been recognized as an important scale in this problem, the so-called ``loitering point'' \citep{Bromm2002}. Thus, as a summary of the effects of the cosmological environment on the thermal evolution we plot the minimum temperature at densities above $100 \, \rm cm^{-3}$ of the collapsing gas in \cref{fig:T_min}. At low halo masses and Lyman-Werner intensities, the $\hd$ cooling mode is active and the minimum temperature can reach as low as tens of Kelvin. At higher halo masses or at intermediate radiation intensities, only $\hto$ cooling is effective with $T_{\rm min} \approx 200 \, \rm K$. Finally, at high $J_{21}$, molecular cooling is shut down and only atomic cooling is possible leading to high gas temperatures of several thousand Kelvin. For some of these halos, the minimum temperature for atomic cooling $\approx 10^4 \, \rm K$ greatly exceeds the virial temperature. In our one-zone model, the gas continues heating adiabatically until this temperature is reached, but in reality the temperature would stall near the virial temperature until the halo grows. We will return to this issue in \cref{sec:final_model}. The physical importance of the temperature minimum will become evident in \cref{sec:cloud_scale}, and the impact on the endpoints of the star formation process will be explicated in \cref{sec:cluster_evo}.

\subsection{Cloud Scale}
\label{sec:cloud_scale}

The characteristic mass of Pop.~III star-forming clouds has long been associated with the temperature minimum of molecular cooled gas around $10^4\,\rm cm^{-3}$, the so-called ``loitering point'' \citep{bromm_forming_1999,Bromm2002,hirano_one_2014,gurian2024zero}. In \cite{Gurian2025}, we argued that this association should not be understood in the sense of the scale on which a density perturbation grows most rapidly \citep[e.g.][]{Larson2005}, but instead in terms of the dynamics after the formation of the first protostar, in the accretion phase. These dynamics have previously been described analytically by similarity solutions \citep{Larson69,Penston1969,Shu77}, giving rise to the characteristic infall rate $\dot M \sim c_s^3/G$, with $c_s$ the sound speed and $G$ the gravitational constant. In a similar vein, we developed a model of the collapse dynamics which accommodates general barotropic equations of state for the collapsing gas. We showed that the features in the density-temperature relationship (controlled by the microphysics of the cooling) naturally produce features in the infall rate during the accretion phase, the cloud-scale infall rate. In the canonical example with $\hto$ cooling, the infall rate drops dramatically after $\sim 1000 \, \rm M_\odot$ (the Jeans mass at the loitering point) has fallen onto the central protostars/protostellar disk \citep[see also][]{BL_acc2004}. 

\cite{Gurian2025}, demonstrated how the chemical-thermal evolution in the primordial gas controls the cloud-scale infall rate. Numerous suites of simulations as well as the analytic model we developed in \cite{liu2024} in turn show that this infall rate is a critical controlling variable in the problem. In particular, for a given infall rate, the total stellar mass in a Pop.~III cluster seems to be relatively robust to the details of the small-scale fragmentation and feedback processes which shape the form of the IMF \citep{Prole2022,Sugimura2023,liu2024}. 
We here briefly review the model of \cite{Gurian2025} while introducing extensions to include turbulent and rotational support. For further details, we refer the reader to \cite{Gurian2025}. 

During the initial pre-stellar collapse, the central region of the cloud is a nearly isothermal core with mass of order the local Bonnor-Ebert mass \citep{Bonnor1956,Ebert55}.  \cite{Gurian2025} extended the Bonnor-Ebert mass to general barotropic equation of states, termed $M_{\rm MBE}$ (for modified Bonnor-Ebert). At a given central density $\rho_c$ during the collapse, interior to $M_{\rm MBE}(r_{\rm MBE}(\rho_c))$ the density profile is approximately hydrostatic, because the sound crossing time is short compared to the cooling time. Adopting spherical symmetry, the equation of hydrostatic equilibrium is: 
\begin{equation}
a_g\equiv\frac{G \left[M_{\rm DM}(r) + M(r)\right]}{r^2} = -\frac{1}{\rho}\frac{dP}{d\rho}\frac{d\rho}{dr},
\label{eq:hse}
\end{equation}
where $M_{\rm DM}$ is the dark matter mass, $M$ the enclosed baryon mass, $\rho$ the gas density, and the pressure $P$ and its derivative are supplied by the effective barotropic equation of state
\begin{equation}
    P_{\rm eff}(n) =  n k_{\rm B} T(n)(1 +{\cal M}^2/3),
\end{equation}
which has been updated from \cite{Gurian2025} to include turbulent support via the turbulent Mach number $\cal M$ \citep[e.g.][]{Li_2021}. The Mach number is assumed to be independent of $n$ for simplicity.

As the collapse proceeds, the central density $\rho_c$ which sets the initial condition for integrating \cref{eq:hse} increases, and $M_{\rm MBE}$ decreases. Meanwhile, outside the core ($r>r_{\rm MBE}$) the density hardly changes over the course of the collapse due to its highly non-homologous nature: the central mass-scale and timescale become extremely small compared to the corresponding peripheral scales. Thus, the sequence of radii and densities at the modified Bonnor-Ebert scale $\big(r_{\rm MBE}(\rho_c), \rho_{\rm MBE}(\rho_c)\big)$ as $\rho_c$ evolves determines the late-time, gravitationally unstable density profile. Qualitatively, the idea is that gravitational instability is induced in the envelope of the cloud by the increased mass concentration from the efficiently cooling, contracting core.   

Then, the velocity profile of the collapsing gas near the time of the first proto-star formation can be estimated by the trajectory of a test particle in this late-time density profile $\rho$ (that is, the density profile when the central number density reaches some maximum density, which we take as $n_{\rm c}=10^{13}\ \rm cm^{-3}$)\footnote{The infall rate at high densities/small masses $\lesssim M_{\rm MBE}(\rho_{\rm max})$ is sensitive to this choice. For $10^{13}\rm cm^{-3}$ and molecular hydrogen cooling,  $M_{\rm MBE} \sim \text{few} \times \, \rm M_{\odot}.$}:
\begin{equation}
    \frac{dv}{dt} = -\beta\left(\frac{G [M_{DM}(r) + M(r)]}{r^2} + \frac{1}{\rho} \frac{dP_{\rm eff}}{d\rho}\frac{d\rho}{dr}-a_c\right),
    \label{eq:vprof}
\end{equation}
where the factor $\beta=1/2$ (which was not included in \cite{Gurian2025}) to calibrate the model against the 1D simulations of \cite{Omukai2010}. This is needed due to the fact that the gas is in reality accelerated over a window of intermediate times and central densities rather than by the late-time density profile. Meanwhile, $a_c$ is the centrifugal acceleration due to rotational support in the cloud. Generally,
\begin{equation}
     a_c =\frac{j^2}{r^3},
\end{equation}
with $j$ the specific angular momentum. We define the cloud-scale rotation parameter 
\begin{equation}
    \lambda^2 = \frac{J^2}{2G M^3 R},
\end{equation}
with $J$ the cloud angular momentum, $M$ the cloud mass, and $R$ the cloud radius. There is in principle some ambiguity in this definition, because it requires choosing an epoch and a definition of the cloud mass and radius. However, in our model $\lambda$ is a free parameter which is not defined in terms of any other model quantities. To obtain the relation between $a_c$ and $\lambda$, we will calibrate the model based on the simulation results of \cite{Sugimura2023}, where the cloud properties are evaluated when the central density is $10^{7} \, \rm cm^{-3}$ and the cloud mass and radius are defined by the scale where the ratio of the enclosed mass to the local Jeans mass attains a maximum (see also \citealt{hirano_one_2014}). 

In simulations, the ratio of centrifugal to gravitational acceleration is often roughly constant over the bulk of the collapsing cloud, but the degree of rotational support declines at lower densities \citep{Stacy_2014a,Sugimura2023,hirano2025dark}. 

Generally,
\begin{equation}
    a_c=a_g\frac{j^2}{GM(r)r},
\end{equation}
so that for a power-law density profile $n \propto r^{-\psi}$, 
\begin{equation}
    a_c \propto a_g n^{(4-\psi)/\psi}. 
\end{equation}
In order to transition from the constant rotational support in the outer part to the constant specific angular momentum regimes in the central region of the collapsing cloud, we make the ansatz
\begin{equation}
a_c =   a_g (\xi\lambda)^2 \left[{1+\left(\frac{n_0}{n}\right)}\right]^{-(4-\psi)/\psi},
\label{eq:a_c}
\end{equation}
where we take $n_0 = 10^3\, \rm cm^{-3}$ as the characteristic cloud density and evaluate $\psi$ by the log-slope of the density profile at $n_0$. Finally, we calibrate $\xi = 4$ against the simulations of \cite{Sugimura2023}. However, at small radii and for large rotation parameters this parameterization can unphysically lead to an outward net force. Thus, we limit $a_c$ such that $\dot v \geq 0$ always. We integrate Eq.~\ref{eq:vprof} inwards from the minimum of 1) the outermost location where $M_{\rm enc} > M_{\rm MBE}$ or, 2) the place where $r/\dot v < 10^9\, \rm yr$, imposing a small ($10^{-2} \, \rm km/s$) inward velocity as the initial condition at this location. A two parameter fit to the centrifugal parameter where both the cutoff scale and $\xi$ are allowed to vary could more realistically capture the dynamics of the rotational support, but this model gives reasonable agreement with the three cases in \cite{Sugimura2023}. 

Finally, the cloud-scale infall rate is given as 
\begin{equation}
    \dot M_{\rm in}(r) = 4\pi r^2 \rho v.
\end{equation}
Through $M(r) = 4 \pi\int dr \,  r^2 \rho$, this equation yields $\dot M_{\rm in}(M),$ which is the rate at which mass is delivered to the central region of the collapsing cloud, as a function of the mass which has already fallen onto the central protostars/disk. This quantity is related to the proto-stellar accretion rate as a function of proto-stellar mass, but the two cannot be immediately conflated because rotational support mediates between the cloud-scale infall rate and the protostellar-accretion rate via a proto-stellar disk (see Section~\ref{sec:cluster_evo} below). 

The modified Bonnor-Ebert mass and the velocity depend on the total mass enclosed, so we adopt throughout this work the dark matter profile used in \cite{Gurian2025} \citep[see also][]{hirano2025dark}:
\begin{equation}
    \rho_{\rm DM}(r) = \frac{\sqrt 2 \rho_s}{\left(\frac{r}{r_s}\right)^{3/2}\left(1+ \frac{r}{r_s}\right)^{1/2}}.
    \label{eq:dmsim}
\end{equation}
We showed in \cite{Gurian2025} that for typical cases the results are sensitive to this choice primarily at masses/radii larger than the cloud scale which do not much affect the final stellar mass: by the critical density of $\sim 10^{3}\,\rm cm^{-3}$ (where collisional de-excitation of molecular hydrogen competes with radiative de-excitation, leading to an increase in the gas temperature with density) the dark matter is subdominant. However, in atomic cooling halos where the gas is closely isothermal, as well as in cases where rotational support reduces the infall rate in the central region, the dark matter density profile can play an important role in regulating the infall. We defer a full exploration of the effects of the dark matter profile on the evolution to future work. 

To sum up, the essential inputs of the cloud-scale model are as follows: a density-temperature relationship (which defines the effective barotropic equation of state), a dark matter density profile (which is relevant only at large radii/masses), a spin parameter, and a turbulent Mach number.

\begin{figure*}
     \centering
    \includegraphics[width=\linewidth]{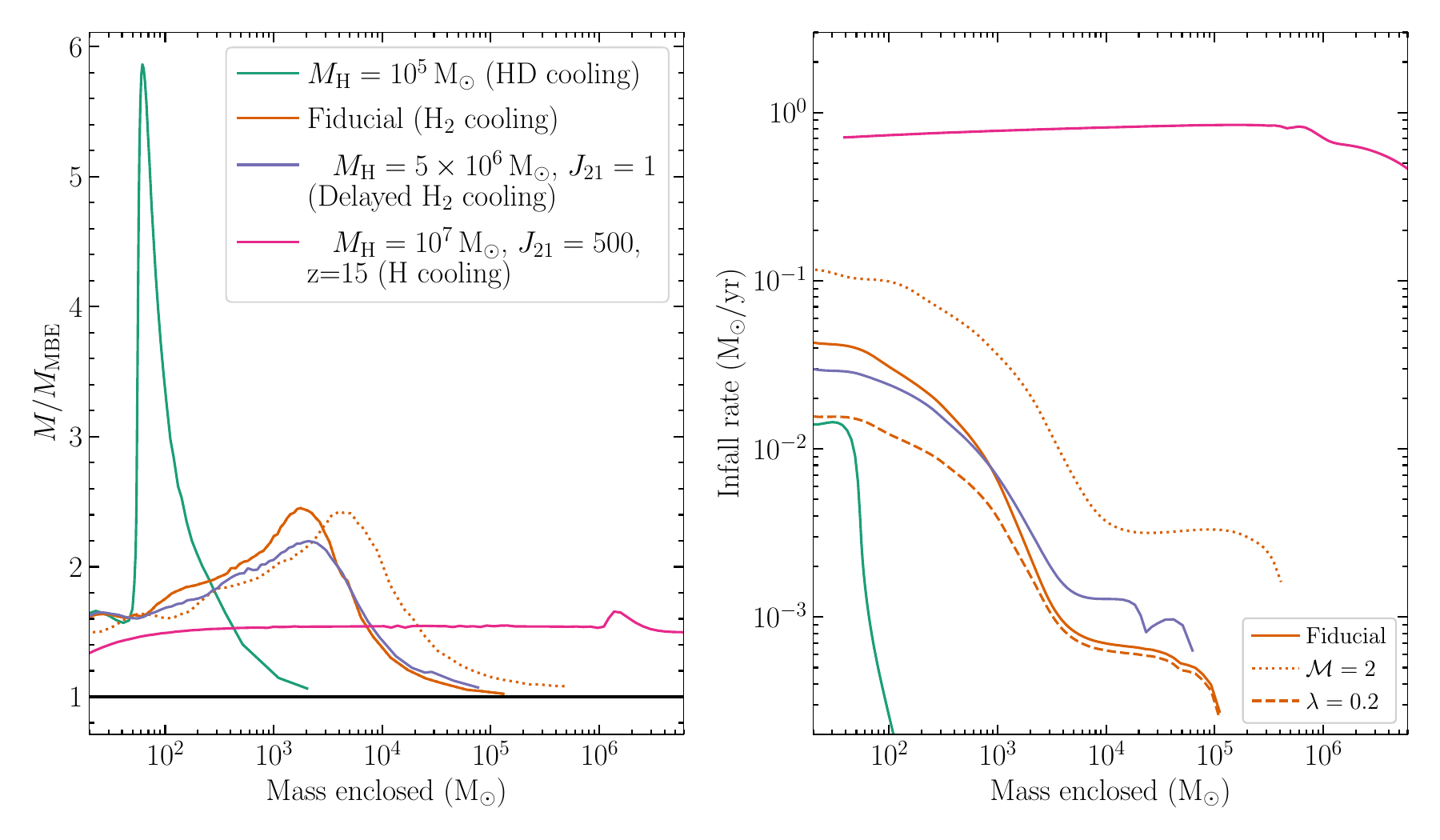}
    \caption{Left: the ratio of the mass enclosed to the modified Bonnor-Ebert mass corresponding to the thermal evolution tracks shown in \cref{fig:n-t_tracks}.    Right: The cloud-scale infall rates for the same cases. For the canonical $\hto$ cooling case, the results of increasing the degree of rotational support to $\lambda = 0.2$ (dashed) and increasing the turbulent Mach number to ${\cal M} = 2$ (dotted) are also shown, although changing the rotation parameter has no effect of $M/M_{\rm MBE}.$ Local minima in the density-temperature relationship map to maxima in $M/M_{\rm MBE}$ and to large $\dot M$.}
    \label{fig:var_jlw_nt}
\end{figure*}

 As an illustration of the range of thermal evolutions and their effects on the cloud-scale infall, we calculate the ratio  $M/M_{\rm MBE}$ and the cloud-scale infall rate for the four cases in \cref{fig:n-t_tracks}. Additionally, for the canonical $\hto$ cooling case we show the evolution for a more turbulent ${\cal M} =2 $ cloud and for a more rotationally supported $\lambda = 0.2$ cloud. The clud-scale parameters are otherwise as in Table~\ref{tab:fid_pars}. The infall rates are systematically higher for the highly turbulent  (${\cal M}=2$) and lower for the fast rotating ($\lambda = 0.2$) cases at fixed thermal evolution. \Cref{fig:var_jlw_nt} shows the sharp drop-off in infall rates at large masses, which can be identified with the temperature minimum in the thermal evolution and with the peak in $M/M_{\rm MBE}$.  Thus, we adopt the highest maximum of $M/M_{\rm MBE}$ as an operational definition of the cloud mass. The dark matter profile we have adopted is based on a sample of low mass mini-halos $\lesssim 10^6 \, \rm M_\odot$. When using this profile for the atomic cooling case, the gas dominates the potential which explains why the infall continues even beyond $\Omega_B M_{\rm H}\sim1.5\times 10^{6} \, \rm M_\odot$ (the baryon fraction times the halo mass). This can occur, for example due to supersonic streaming, although in that case the turbulence of the cloud would likely be enhanced. 

The model parameters influence the evolution in two distinct ways: through the infall rate within the cloud and through the mass of the cloud itself (or, equivalently, the scale at which the infall rate drops precipitously). The cloud mass is controlled principally by the gas thermal evolution, and so is sensitive primarily to $J_{21}$, $f$ (which controls especially the transition to low cloud masses with efficient $\hd$ cooling), and the redshift $z$ (which sets the CMB temperature floor, relevant in the $\hd$ mode). On the other hand, the Mach number and especially the rotation parameter principally alter the infall rate \textit{within} the cloud. 

These parameters are in reality not independent. For example, a large delay in the collapse leads to growth of the halo and corresponding changes in the dark matter profile which can alter the infall rate at large enclosed mass. In particular, the dark matter density profile we adopt is likely not appropriate for the atomic cooling halos. Further, the delay parameter $f$ is largely determined by other parameters, as discussed in the previous section. We neglect any dependence of $\lambda$ and $\cal M$ on other variables. 

\begin{figure*}
    \centering
    \includegraphics[width=\linewidth]{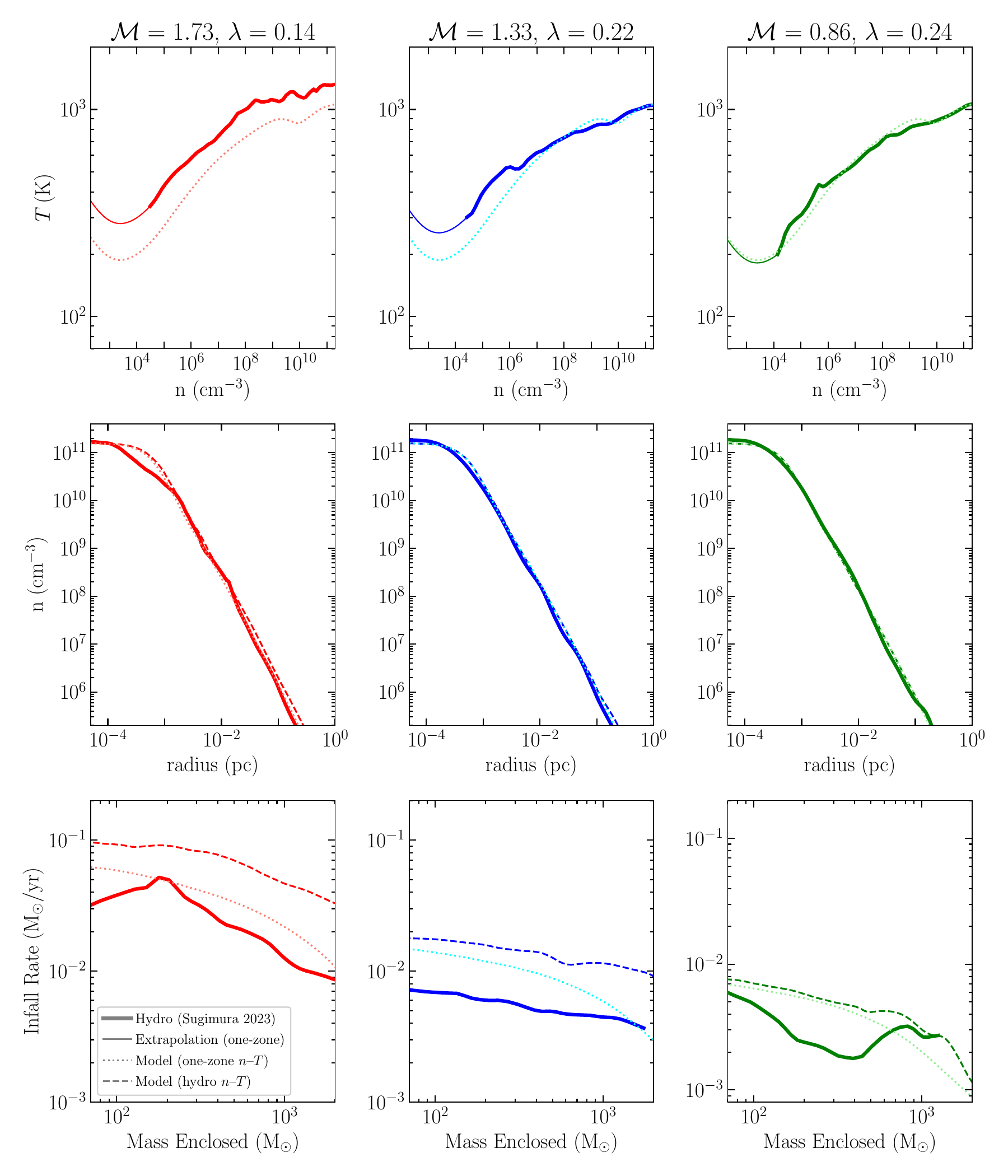}
    \caption{The simulations of \protect\cite{Sugimura2023} compared with the model results for three clouds labeled by the turbulent Mach number and rotation parameter. Left, center, and right correspond respectively to the high, intermediate, and low $\dot M$ cases of \protect\cite{Sugimura2023}. Top row: the density-temperature relationships from the simulations of \protect\cite{Sugimura2023} (thick, solid lines), extrapolated to lower densities using a one-zone model (thin, solid lines), and from a one-zone calculation (thin, dotted lines). Middle row: the density profile from simulations of \protect\cite{Sugimura2023} (thick, solid) and from the cloud model with the simulation $n$-$T$ relationship (dashed) and the one-zone $n$-$T$ relationship (dotted). Bottom row: the cloud-scale infall rate as a function of cumulative collapsed mass, for the same cases. }
    \label{fig:cloud_compare}
\end{figure*}

In Fig.~\ref{fig:cloud_compare} we compare the model with the three cases from the simulations of \cite{Sugimura2023} (low, intermediate, and high $\dot M$), both using the $n$-$T$ relation produced by the 3D simulation as an input, and using a one-zone $n$-$T$ relationship produced by solving the chemical-thermal network with the fiducial redshift, halo mass and delay parameter of \cref{tab:fid_pars}.  We adopt the rotation parameters reported in \cite{Sugimura2023}, while the turbulent Mach number is a mass-weighted average from the simulation data when the central density is $\sim 10^{11}\, \rm cm^{-3}$. The lowest density in the $n$-$T$ relationship from \cite{Sugimura2023} is $10^5\, \rm cm^{-3}$, at which point the gas is already gravitationally unstable, $M_{\rm enc} > M_{\rm MBE}$ and has a non-negligible inwards velocity. Thus, in those cases we extrapolate the density-temperature relationship to low density using the one-zone model, rescaled to ensure continuity with the simulations results. Further, we comment that in the cosmological simulation of \cite{hirano_one_2014} (from which the clouds of \cite{Sugimura2023} are initialized) $\hd$ cooling is active in the low $\dot M$ (right hand column of Fig.~\ref{fig:cloud_compare}) case, but \cite{Sugimura2023} do not include $\hd$ which leads to an upward adjustment of the temperature at $n \sim 10^5\ \rm cm^{-3}$ between the initial condition and the final state. This abrupt change in cooling rates cannot be accommodated within our model. Still, in all cases the model predicts the infall rate to within a factor of $3$ accuracy over the full evolution.

\subsection{Disk Scale}
\label{sec:cluster_evo}
In \cite{liu2024}, we developed an algebraic model describing the growth of Pop.~III stars in a protostellar accretion disk, under an assumed self-similar growth. Here, we develop this into a model describing the time evolution of the proto-stellar mass function. We begin by briefly summarizing the model of \cite{liu2024}. In the fiducial case, the number of protostars was held fixed while the upper and lower limits of the mass function evolved due to accretion. The growth of the cluster terminates as soon as one of the following conditions was met:
\begin{itemize}
    \item The most massive star in the cluster reaches the general relativistic instability, or
    \item the shortest-lived (most massive) star in the cluster reaches the end of its lifetime, or 
    \item The supply of gravitationally unstable gas to the protostars is depleted (by being accreted onto protostars and by photoevaporation). 
\end{itemize}
Here, we develop these same physical ingredients into a model of the time-evolution of the growing (proto-)star cluster fed by the protostellar disk. Specifically, we write down the following system of ordinary differential equations describing the cluster and disk evolution:
\begin{align}
     \dot M_d &= \dot{M}_{\rm in} - \dot M_* - {\dot M_{\rm pe}} \label{eq:mdotd}\\
  \dot M_{\rm pe}&= 0.015 {\, \rm M_\odot yr^{-1}}\left[\frac{\int_{m_1}^{m_2}  \, \frac{dN}{dm}  \dot q_{\rm ion}(m)dm}{10^{52 }\, \rm s^{-1}}\right]^{1/2}\left(\frac{r_d}{10^4 \, \rm AU}\right)^{1/2} \label{eq:mdotpe}\\
      \dot M_* &= \frac{\epsilon M_d}{M_d + M_\star} \dot M_{\rm in}, \label{eq:mdotstar}
\end{align}
where $M_d$ is the disk mass, $\dot{M}_{\rm in}$ the cloud-scale infall rate, describing the rate at which the gravitationally unstable cloud supplies mass to the disk, $ \dot{M_\star}$ the total accretion rate onto protostars, $\dot M_{\rm pe}$ the disk photoevaporation rate computed as in \cite{liu2024} (see also \citealt{Toyouchi_2022,Tanaka_2013}), given the mass function of protostars $dN/dm$ (that is, $\int_{m_1}^{m_2}m(dN/dm)dm=M_\star$), and $\dot{q}_{\rm ion}(m)$ the production rate of ionizing photons as a function of stellar mass\footnote{In fact, $\dot{q}_{\rm ion}$ depends on both the stellar mass $m$ and the accretion rate $\dot m$. In our case, $\dot m$ is derived from $m$ given the total accretion rate $\dot M_\star$ using a fragmentation model detailed in Sec.~\ref{sec:lowerlimit} and Appendix~\ref{app:mdot}.}, $r_d$ is the disk radius, and $\epsilon$ is the equilibrium ratio of stellar to disk mass. The disk radius is given by the scaling derived in \cite{liu2024}
\begin{equation}
    r_d= \min\left(\left[10^{5\beta}M_{\rm in}/(400\ {\rm M_\odot})\right]^{\delta/\beta}\ {\rm AU},\, 10^5 \, {\rm AU}\right),
\end{equation}
with  $M_{\rm in}$ the total infall-mass, $\beta=4-3\gamma_{\rm eff}$, and $\delta=2-\gamma_{\rm eff}$. We adopt $\gamma_{\rm eff} = 1.09$ as the typical effective adiabatic index near the disk edge. Actually, $\gamma_{\rm eff}(M_{\rm in})$ is determined by the chemical-thermal evolution and density profile, but \cite{liu2024} showed that this parameter does not strongly affect the results, so we take $\gamma_{\rm eff}$ fixed at a typical value for convenience. 

Eq.~\ref{eq:mdotstar} is an ansatz which describes the factor-of-few difference between the cloud-scale infall rate and the proto-stellar accretion rate due to rotational support that funnels a fraction of the infalling gas to disk growth \citep{Kimura2021}. This means that when $\dot{M}_{\rm pe}$ is small in the early stage, the system will converge to a (steady) state where the stellar-to-disk mass ratio is $\epsilon$ and the corresponding mass fraction of protostars in the central collapsed (stars+disk) system is $\eta\equiv \epsilon/(1+\epsilon)$. Hydrodynamic simulations and analytical models normally find that the stellar-to-disk ratio increases over time, and we typically have $\epsilon\sim 1/3-3/2$ and $\eta\sim 0.25-0.6$ on average \citep{Sakurai2016,Kimura2021,Toyouchi_2022,Sharda2025}. Following \citet{liu2024}, we adopt $\epsilon=1$ (i.e. $\eta=0.5$) as a constant for simplicity, based on \citet{Sakurai2016} and \citet{Toyouchi_2022}. On the other hand, when $\dot{M}_{\rm pe}$ is large, the growth of protostars will naturally end when the star-forming disk is evaporated, i.e. $M_{d}\rightarrow 0$. 

\subsubsection{Single Star Case}
\label{sec:singlestar}
As a first demonstration, we solve \cref{eq:mdotd,eq:mdotstar,eq:mdotpe} considering the single star case with $dN/dm = \delta(M-M_\star)$, where $\delta$ is the Dirac delta. \cite{Liu2021binary} found a ``universal'' power-law infall rate which holds in the early accretion stage, where the gas has an effective adiabatic index $\gamma_{\rm eff} \approx 1.09$, across numerous simulations conducted by different groups. In \cite{liu2024} we supplied $\dot M_{\rm in}$ by rescaling this universal (scale-free) infall rate by an overall factor, to study the dependence of the stellar mass on the infall rate. However, \cite{Gurian2025} showed that the ``loitering point'', at which the temperature of the gas bottoms out and the effective adiabatic index becomes softer, induces a cutoff in the infall rate at large masses (see \cref{fig:var_jlw_nt}). As shown in the previous section, both the cutoff scale and the amplitude of the infall depend on the collapse parameters. Thus, there is not a one-to-one mapping between the infall rate at some particular scale and the final stellar mass. Further, the distributions and covariances of the collapse parameters are highly non-trivial, and should in principle be supplied by cosmological data (e.g.~simulations or merger trees). In order to calibrate the model against 3D simulations while bypassing these difficulties, we generate a fiducial $\hto$ cooling infall function $\dot M_{\rm \hto}(m)$ using the parameters shown in \cref{tab:fid_pars}. Then, we rescale this fiducial infall law as 
\begin{equation}
    \dot M_{\rm scaled}(M) = a\dot M_{\hto}(M/a).
\end{equation}
Typically, high mass infall rates are realized in atomic cooling halos, so we also calculate $\dot M_{\rH}(M)$ where the parameters are the same except that $J_{21}=10^4$, and rescale this infall rate. For consistency with \cite{liu2024}, we label the fiducial infall-rate tracks by the infall rate at the place where the density is $n=10 ^6 \, \rm cm^{-3}$, and the rescaled rates by the appropriately rescaled fiducial infall rate: $\dot M_{\rm in,6} = a\dot M_{\rm in,6\, fid} $. We integrate the evolution equations with the initial condition $M_* = M_d =  a \, \rm M_\odot$ (i.e.~$1 \, \rm M_\odot$ in the fiducial case). 

\begin{figure}
    \centering
    \includegraphics[width=\linewidth]{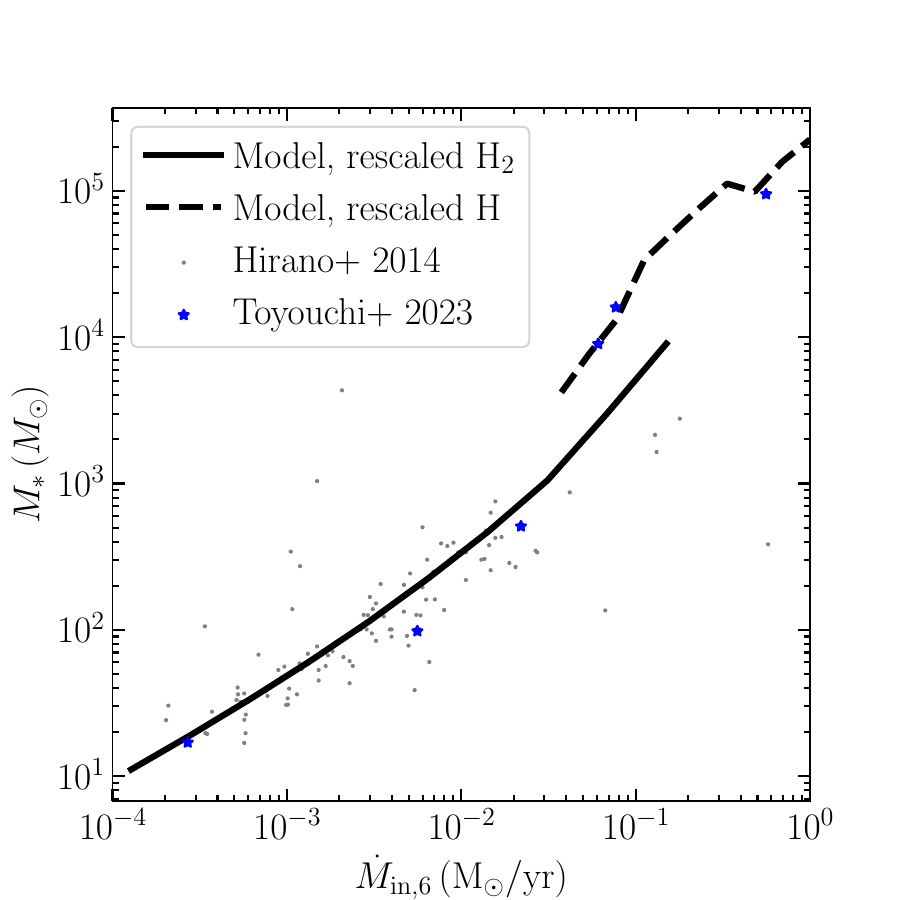}
    \caption{ The stellar mass in the single star case at the time of disk photo-evaporation, the end of the stellar lifetime, or the onset of general-relativistic instability, for the rescaled infall rates for the molecular cooling case (solid) and the atomic cooling case (dashed). Also shown are the simulation samples of \protect\cite{hirano_one_2014} and \protect\cite{Toyouchi_2022}.}
    \label{fig:res_onestar}
\end{figure}

We terminate the integration once $M_d=0$, or $\dot M_*$ exceeds the general relativistic instability threshold at $M_*$, or the elapsed time exceeds the lifetime of $M_*$, calculated as in \cite{liu2024}. The results are shown in \cref{fig:res_onestar}, compared with the simulation samples of \cite{hirano_one_2014} and \cite{Toyouchi_2022}. In both simulation suites, fragmentation is artificially suppressed (by azimuthal averaging in the former case and low resolution away from the central region in the latter), so that the single star comparison is appropriate. Although this rescaling is somewhat ad-hoc, it is able to reproduce the simulation trends quite well. Note that the minimum stellar lifetime was calibrated in \cite{liu2024} to ensure monotonicity in stellar mass as a function of infall rate: in this updated model the transition from ``lifetime limited'' to ''GR instability limited'' produces a kink around $\dot M_{\rm in} =0.4\,\rm M_\odot/yr$. These results illustrate how the interplay between the available supply of gravitationally unstable gas, the disk photo-evaporation, and stellar evolution dictates the final stellar mass in the absence of fragmentation. Interestingly, we already see the limitations of the one-parameter rescaling, in that the rescaled $\hto$ track better fits the high infall rate molecular cooling clouds of \cite{hirano_one_2014}, while the rescaled $\rm H$ track better matches the nearly isothermal atomic cooling clouds of \cite{Toyouchi_2022}.  

\subsubsection{Fragmentation: Fixed Lower Cutoff}
\label{sec:lowerlimit}

To generalize beyond the single star case, we assume that the mass function is a pure power law with a fixed lower limit and time-evolving upper limit, 
\begin{equation}
    \frac{dN}{dm} = \int_{m_1}^{m_2(t)}c(t) m^\alpha.
\end{equation}

Physically, this is the assumption that the mass function at intermediate masses evolves in a scale free way, while the lower limit is set by some time-independent physics in the disk and the upper limit is dictated by the growth rate of the most massive star. This scenario is plausible if the characteristic lower limit of the mass function is set by the dynamical/thermal conditions in the molecular disk (which vary only weakly over the course of the proto-stellar accretion) or by the cooling physics of the gas (via the opacity limit). To some extent this is the opposite limit of the fiducial model of \cite{liu2024}, where an early epoch of fragmentation is assumed to produce a fixed number of protostars which together undergo self-similar growth.  The assumption of  a single power law form is reasonable especially for a top-heavy mass function as is expected for Pop.~III stars, where the mass budget can be dominated by massive stars described by a power law distribution \citep[e.g.][]{hirano_one_2014,Stacy2016,Prole2022}.
We do not consider individual realizations of the fragmentation spectrum but instead consider the mass function, accretion rates, together with other variables of the problem, as stochastic quantities and model the evolution of their expectation values. 

\Cref{eq:mdotpe} depends on the proto-stellar accretion rate (as a function of proto-stellar mass) because at high accretion rate $\gtrsim 0.02\,\rm M_\odot/yr$ the protostars enter a ``bloating'' cool supergiant phase and hardly emit ionizing photons \citep{hosokawa_rapidly_2012}. Subject to the time-dependent constraints on the total mass and total accretion rate (\cref{eq:mdotstar}) and the assumed power law form of the mass function, this proto-stellar accretion rate $\dot m(m)$ is derived in Appendix~\ref{app:mdot}. Conceptually, the picture is similar to the standard ``competitive accretion'' scenario, where a power-law accretion rate (as a function of mass) naturally produces a power-law mass function. In our case, the algebra is more complicated due to the fixed lower mass limit and the ``global'' constraints imposed through $\dot{M}_{*}$.

In this derivation, $\dot m_2$ (the time derivative of the upper limit of the mass function) remains undetermined. One simple ansatz is 
\begin{equation}
    \dot m_2 = \chi \dot M_*,
\end{equation}
with $\chi$ a constant. Motivated by \cite{Greif2012}, we adopt $\chi = 0.6$ as the fiducial choice. An alternative parameterization derived in Appendix~\ref{app:mdot} relates $\dot m_2$ to the rate at which protostars are produced at $m_1$, $\dot n$.

The constraint on the total mass also determines the time derivative of $c$, derived in Appendix~\ref{app:mdot} as:

\begin{equation}
    \dot c(t)= \left(\frac{\dot M_*}{M_*}-\frac{m_2^{1+\alpha}c(t)}{M_*}\frac{\partial m_2}{\partial t}\right)c(t).
    \label{eq:cdot}
\end{equation}

Given $\dot m_2$, we integrate \cref{eq:cdot} simultaneously with \cref{eq:mdotd,eq:mdotpe,eq:mdotstar}. The integration is terminated once $M_d=0$, or the lifetime of the most massive star in the cluster is exceeded, or the accretion onto the most massive star exceeds the general relativistic instability limit. For small to moderate $N$ the occupation number of the mass function will be less than unity near $m_2$. A characteristic mass of the most massive star in the cluster $m_u$ which we adopt is:
\begin{equation}
    \int_{m_u}^{m_2}\frac{dN}{dm}dm= 1,
\end{equation}
or 
\begin{equation}
m_u=\left(m_2^{\alpha +1}-\frac{\alpha
   +1}{c}\right)^{\frac{1}{\alpha +1}}.
\end{equation}

Although this model is highly simplistic, it is a self-consistent toy model which captures the interplay between proto-stellar growth, radiative feedback, and fragmentation. Further, the scenario of vigorous fragmentation at small masses which is revealed by high resolution simulations \citep[e.g.][]{susa_merge_2019,Prole2022} cannot be self-consistently evolved through the shutdown of accretion with current computational resources. This model enables exploration of the scenario where the distribution of Pop.~III stellar masses within a cluster is very wide. 

As a first application, we calculate the final stellar properties for the four cases shown in \cref{fig:n-t_tracks} and \cref{fig:var_jlw_nt}, using the fiducial fragmentation parameters (\cref{tab:fid_pars}). Note that the range of stellar and cluster masses is very wide, and that the total stellar mass is a factor of a few small than the cloud masses (the peaks of $M/M_{\rm MBE}$ in \cref{fig:var_jlw_nt}).

\begin{table*}
\centering
\caption{Stellar properties calculated using the fiducial disk fragmentation model for the cases shown in \cref{fig:var_jlw_nt}.}
\label{tab:cooling_comparison}
\renewcommand{\arraystretch}{1.5}
\begin{tabular}{lcccccc}
\hline
 & \textbf{$\hd$ cooling} &\textbf{\begin{tabular}[c]{@{}c@{}}$\hto$ cooling \\fiducial\end{tabular}} & \textbf{\begin{tabular}[c]{@{}c@{}}$\hto$ cooling \\ $\lambda=0.2$\end{tabular}} & \textbf{\begin{tabular}[c]{@{}c@{}}$\hto$ cooling\\  ${\cal M }=2$\end{tabular}} & \textbf{\begin{tabular}[c]{@{}c@{}}  H$_{2}$ cooling\\ $J_{21}=1$\end{tabular}} & \textbf{H cooling} \\ \hline
Stellar mass $M_*$ ($\rm M_\odot$)       & 59 & 480 & 370 & 1200  & 490 &$ 6.5\times 10^5 $ \\ 
Most massive star $m_u$ ($\rm M_\odot$)     & 27 &  240 & 180 & 610 & 240 & $3.7 \times 10^5 $ \\ 

\end{tabular}
\label{tab:nt_final}
\end{table*}

\begin{figure*}
    \centering
    \includegraphics[width=\linewidth]{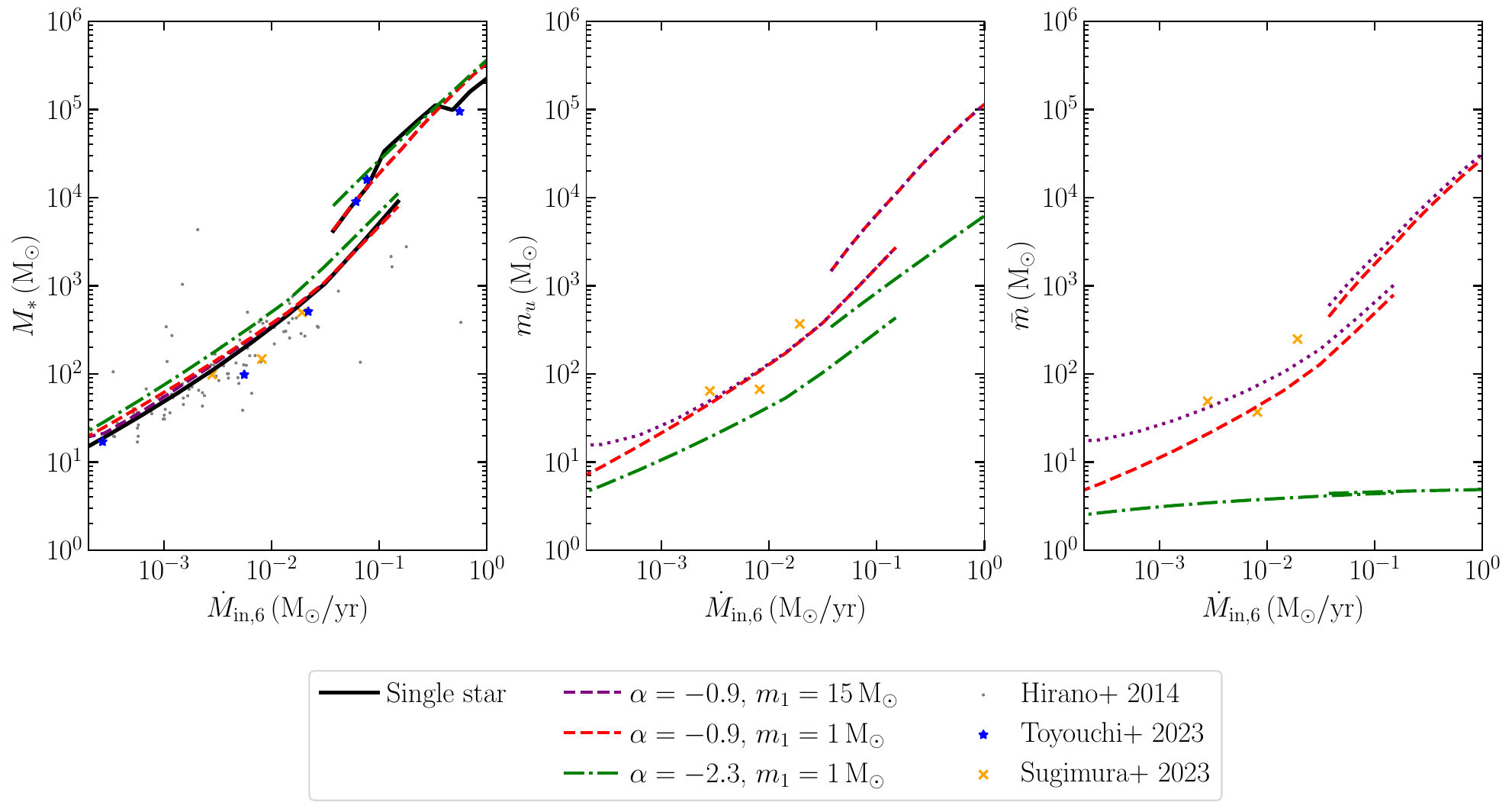}
    \caption{The total stellar mass (left), most massive star (center), and mean stellar mass (right) from the simulations of \protect\cite{hirano_one_2014, Toyouchi_2022,Sugimura2023} and the model of this work in the case with a single star, and with power law mass function $\alpha=-0.9$, and $\alpha=-2.3$. For the case with $\alpha=-0.9$, we show both $m_1=1\,\rm M_\odot$ and $m_1 = 15 \, \rm M_\odot$. The discontinuities in the model results are due to switching from the rescaled $\hto$ infall rate to the rescaled $\rH$ infall rate. Only the simulations of \protect\cite{Sugimura2023} resolve the disk fragmentation so that $m_u$ and $\bar m$ can be meaningfully calculated.}
    \label{fig:alpha_compare}
\end{figure*}

 Now, we apply the model to study the dependence of the total stellar mass, characteristic upper mass $m_u$, and mean mass $\bar m$ on the fragmentation parameters. To this end, we again adopt the rescaled infall rates of \cref{sec:singlestar}, taking $m_1= 1 \, \rm M_\odot$ and the initial condition $M_{*,0}=M_{d,0}=1.1m_1$, with $m_{2}$ chosen so that $\int_{m_1}^{m_2} dm \frac{dN}{dm}=1$, implying that the expected number of stars at the initial time is unity. We consider two power law mass functions, top heavy $\alpha = -0.9$ and Salpeter-like $\alpha = -2.3$. The initial time is $t_0 = 100 / a\, \rm yr$, motivated by the $\sim 0.01 \, \rm M_\odot /yr$ initial infall rates for the $\hto$ cooling case. In \cref{fig:alpha_compare}, we compare these cases with the simulations of \cite{hirano_one_2014,Toyouchi_2022,Sugimura2023}\footnote{As in \cite{liu2024} the infall rates of \cite{Sugimura2023} are multiplied by a factor of $3.7$ because in that work the infall rate is calculated when the central density is $10^7\,\rm cm^{-3}$ \citep[compared with $10^{12}\,\rm cm^{-3}$ in][]{hirano_one_2014} and the model results, and the infall velocity increases over the course of the initial collapse before proto-star formation.}, plotting the total stellar mass, $m_u$  and the mean stellar mass $\bar m$. The total stellar mass is quite insensitive to the degree of fragmentation, as has been previously pointed out \citep[e.g.][]{Prole2022, Sugimura2023}.
In the model results, including fragmentation modestly boosts the star formation efficiency (SFE) due to the less efficient ionizing photon production by low-mass stars. In contrast, in \cite{Sugimura2023} only a handful of massive stars formed, and there is no clear difference in the SFE (recall though the ambiguity in the definition of $\dot M_{\rm in}$). The case of a top-heavy mass function $\alpha = -0.9$ with a high lower mass limit $m_1 = 15 \, \rm M_\odot$ agrees fairly well with the results of \cite{Sugimura2023}. This may be a consequence of the fact that the relatively low resolution in that work inhibits fragmentation on small scales. One interesting observation is that even for the Salpeter mass function, at high infall rates $m_u >1000\,\rm M_\odot$. In present day star formation, the IMF has a cutoff at high masses \citep{Figer_2005}. Simulations have not yet converged on the question of whether the Pop.~III IMF has a similar cutoff \citep[though see][]{Sharda2025}.

Similarly, the disk model can be applied to study the lower limit of the mass function and the growth of the number of protostars over time. In the preceding analysis, the lower limit $m_1$ is not necessarily the mass at which protostars form, but rather a scale at which the evolution is assumed to become self-similar (and at which the ionizing photon production becomes non-negligible). We adopted $m_1=1\, \rm M_\odot$ as our fiducial choice, motivated by the fact that a Pop.~III star $<0.8 \, \rm M_\odot$ could survive to the present day \citep[e.g.,][]{Hartwig2015}. 

Physically, though, protostars are expected to form near the opacity-limited Jeans mass \citep{Rees1976,Yoshida08}. This is where the equation of state of the collapsing gas becomes nearly adiabatic at $\sim 10^{19} \, \rm cm^{-3}$, with an associated Jeans mass $\sim 10^{-2} \, \rm M_\odot$ \citep[for a discussion of the environment-dependence in metal-free gas, see][]{Becerra2018}. Simulations which resolve this density are limited by the high computational cost to evolving only a few years of physical time \citep{Greif2012}. Further, achieving convergence in the fragmentation cascades with respect to the simulation resolution is notoriously problematic \citep{Prole2022}. Higher resolution simulations consistently find fragmentation at ever smaller scales, exactly as expected for opacity-limited fragmentation. 

However, there is some evidence for a self-similar behavior in the fragmentation cascade all the way down to the opacity limit. \cite{susa_merge_2019} showed that in the early stage of the evolution, the number of fragments scales predictably with the resolution limit as 
\begin{equation}
N_* = 3  \left[\left(\frac{t}{\rm yr}\right) \sqrt{\frac{n_{\rm max}}{10^{19}\, \rm cm^{-3} }}\right]^{0.3},
\label{eq:nstar}
\end{equation}
where $n_{\rm max}$ is the resolution limit. 

\begin{figure}
    \centering
    \includegraphics[width=\linewidth]{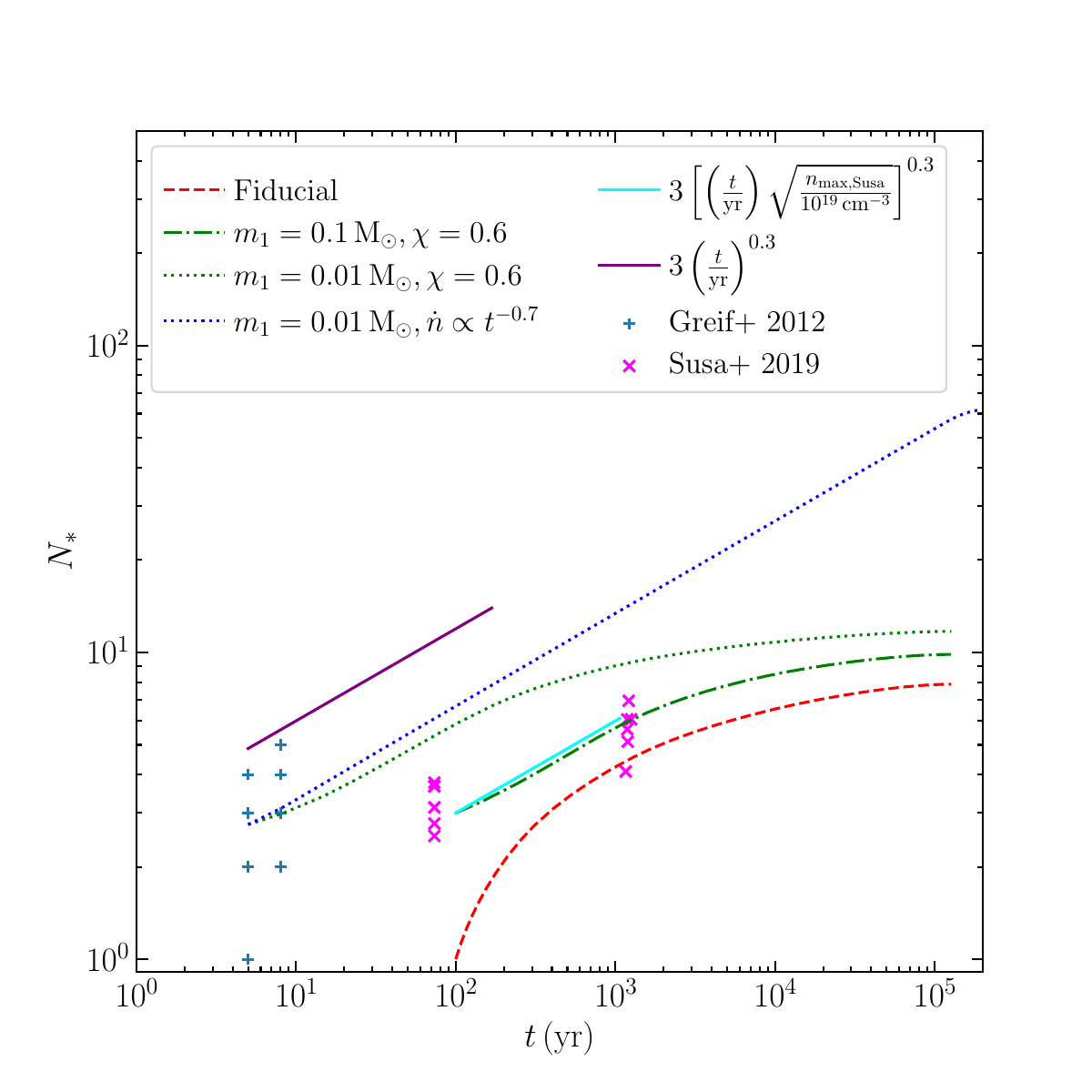}
    \caption{The number of stars over time in the fiducial model with $m_1 = 1 \, \rm M_\odot$ (black, dashed), and model parameters chosen to match \protect\cite{susa_merge_2019} (green, dot-dashed) and \protect\cite{Greif2012} (red, dotted) and \protect\cite{Greif2012} with the injection rate $\dot N \propto t^{-0.7}$ (blue, dotted). The actual number of fragments in the simulations of \protect\cite{Greif2012} and \protect\cite{susa_merge_2019} are also shown (markers). Finally, the scaling relation of \protect\cite{susa_merge_2019} at the adiabatic limit (corresponding to \protect\cite{Greif2012}, purple) and the maximum density in \protect\cite{susa_merge_2019} (cyan).}
    \label{fig:nstar}
\end{figure}

We can apply our fragmentation model to the scenario of self-similar fragmentation down to the opacity limit by extending the one-zone model to $10^{19} \, \rm cm^{-3}$. This resolves the infall rate down to the first $\lesssim 0.01 \, \rm M_\odot$ accreted. We consider three new cases, for which $\alpha = -0.9$. First, we set $m_1=0.1 \, \rm M_\odot$, which matches the resolution limit of \cite{susa_merge_2019}, adopting $t_0=100\, \rm yr$, $M_{*,0} = 7 \, \rm M_\odot$ and $N_{*,0}=3$ to match their findings. Second, we adopt $m_1=0.01 \, \rm M_\odot$, and \citep[matching][]{Greif2012} $t_0=5 \, \rm yr$, and $M_{\star,0}=0.7\, \rm M_\odot$. In this case, $m_1$ truly corresponds to the opacity limit. Finally, we take the same opacity-limited setup but specify $\dot m_2$ by $\dot N \propto t^{-0.7}$, using \cref{eq:m2ofndot}. This is consistent with the scaling \cref{eq:nstar}\footnote{At the initial time, \cref{eq:m2ofndot} gives $\dot m_2 <0$ by a small amount, so we use $\min(0, \dot m_2(\dot N)$.} The results are shown in \cref{fig:nstar}, along with simulation results from \cite{susa_merge_2019} (which had a resolution limit of $10^{15} \, \rm cm^{-3}$) and from \cite{Greif2012}. The scaling relation \cref{eq:nstar} is also shown, both at the true opacity limit and at the resolution limit of \cite{susa_merge_2019}. In the case of \cite{susa_merge_2019}, the scaling relation is plotted over the time span of the simulation. For the case at the adiabatic limit, the scaling relation is plotted over the range of scaled times (which is the physical time rescaled by $\sqrt{n_{\max}/n_{\rm ad}}$) cataloged from numerous independent simulations by \cite{susa_merge_2019}. 

The early evolution is roughly power law and consistent with simulations in both the scenario where the fragmentation is driven by a specified $\dot m_2$ and the scenario where it is driven by $\dot N$ (although in the former case the index of this power law depends on $\chi$). However, in the $\dot m_2 =\chi \dot M_*$ case, the fragmentation is eventually choked off by accretion onto the massive protostars. Although they are described by the same model, these are physically quite distinct scenarios. The evolution of the stellar mass function is dictated either by the growth of the upper mass limit following the cloud-scale infall rate or by the rate of fragmentation set by the conditions in the disk.  This model highlights the tension between these two physically plausible cluster evolution scenarios in the context of a global accretion rate constraint. The distinction between these two scenarios is astrophysically relevant especially to the question of the seeding of massive black holes by Pop.~III stellar remnants in atomic cooling halos. Simulations do not yet provide a clear answer to this question, with some finding that ``super-competitive accretion'' \citep{Chon_2020} or runaway collisions \citep{Reinoso_2025} lead to heavy seeds $\gtrsim 10^4 \, \rm M_\odot$, while others suggest that fragmentation-induced starvation limits the growth of massive stars \citep{Prole2022}.

Despite these caveats, it is tempting at this point to declare victory: a model based on a disk which is scale free except for $m_1$ (since $\dot M_{in}$ is nearly scale-free during the early evolution) can reproduce the trends observed in diverse simulations, and span the range of outcomes where these simulations still disagree. Thus, one might hope that a truly analytic description of the fragmentation cascade in the disk could be developed. However, this scenario completely fails to explain the apparent absence of low mass Pop.~III stars which could survive to the present day. It is this observational fact which motivated the fiducial choice $m_1= 1 \, \rm M_\odot$. From this perspective, the absence of surviving Pop.~III stars is a true mystery \citep[see also][]{Frebel2009}. 

\subsection{Unified Model: Cosmological Dependence of Pop.~III Star Formation}
\label{sec:final_model}
The tools developed in \cref{sec:haloscale,sec:cloud_scale,sec:cluster_evo} enable prediction of the properties of a newly formed Pop.~III cluster from the cosmological environmental variables (redshift, halo mass, and radiation field) for given cloud properties (turbulent Mach number and spin parameter) and fragmentation properties. To develop the model, we have so far focused on particular cases of interest (e.g.~\cref{fig:n-t_tracks,fig:var_jlw_nt}) or have made simplifying parameterizations to study the range of outcomes (as in \cref{sec:singlestar,sec:lowerlimit}). Now, we are nearly ready to return to a full cosmological parameter exploration, as alluded to in the calculation of the minimum temperature of the collapsing gas in \cref{sec:haloscale} (\cref{fig:T_min}). 

Yet, the preceding machinery is focused on answering the question of how Pop.~III star formation proceeds, given particular (cosmological) initial conditions. Before carrying out this calculation, we turn to the issue of which halos are able to produce Pop.~III stars. The basic physical answer is simple: any halo which becomes Rees-Ostriker unstable in the sense of \cref{sec:haloscale} and has not previously formed stars will form Pop.~III stars within approximately a dynamical time. The halo-scale model in this work could be applied to this question by using halo merger trees to track the early chemical-thermal evolution and assess which halos remain pristine at a given redshift, similar to \cite{Li_2021}. 

As a first step in this direction, we make simple arguments to bound the region where Pop.~III star formation is likely to occur. First, the lower limit for star formation (often termed the critical halo mass) has been studied extensively using analytic methods and simulations \citep{Tegmark1997,Machacek2001,Yoshida03,Kulkarni_2021,Schauer_2021,Nebrin:2023yzm}. We adopt the fitting function for $M_{\rm crit}$ from \cite{Kulkarni_2021}. In that work, $M_{\rm crit}$ was defined as the mass above which half of halos have undergone gas collapse, but by inspection of their figures~1 and 6, gas collapse is very rare in halos with masses below $M_{\rm crit}/2$, which we adopt as a lower limit for the mass of Pop.~III star-forming halos.
Next, we argue that if a given halo can become Rees-Ostriker unstable on a timescale much shorter than the Hubble time, it is likely that a lower mass progenitor has already formed stars that led to metal enrichment. We correspondingly exclude Pop.~III star formation in regions with $t_{\rm RO} < t_{H}/2$, where $t_{H}$ is the Hubble time. 

At high redshift the halo mass function imposes a constraint on star formation in more massive halos. It is thus also informative to consider the abundance of halos $n(>M_H)$, where $n(>M_H)$ is  the integrated number of halos with mass above $M_H$. We calculate the halo mass function using the Press-Schechter formalism \citep{Press1974}, via the package Colossus \citep{Diemer18}. 

Finally, the low redshift boundary of Pop.~III star formation is a highly observationally relevant open question which depends on the survival of pockets of pristine gas around massive halos \citep{Mebane18,Liu2020sf,Venditti2023, Venditti2024,Fujimoto25,storck2025}.  We do not attempt to place a boundary in this region. 

\begin{figure}
    \centering
    \includegraphics[width=1.1\linewidth]{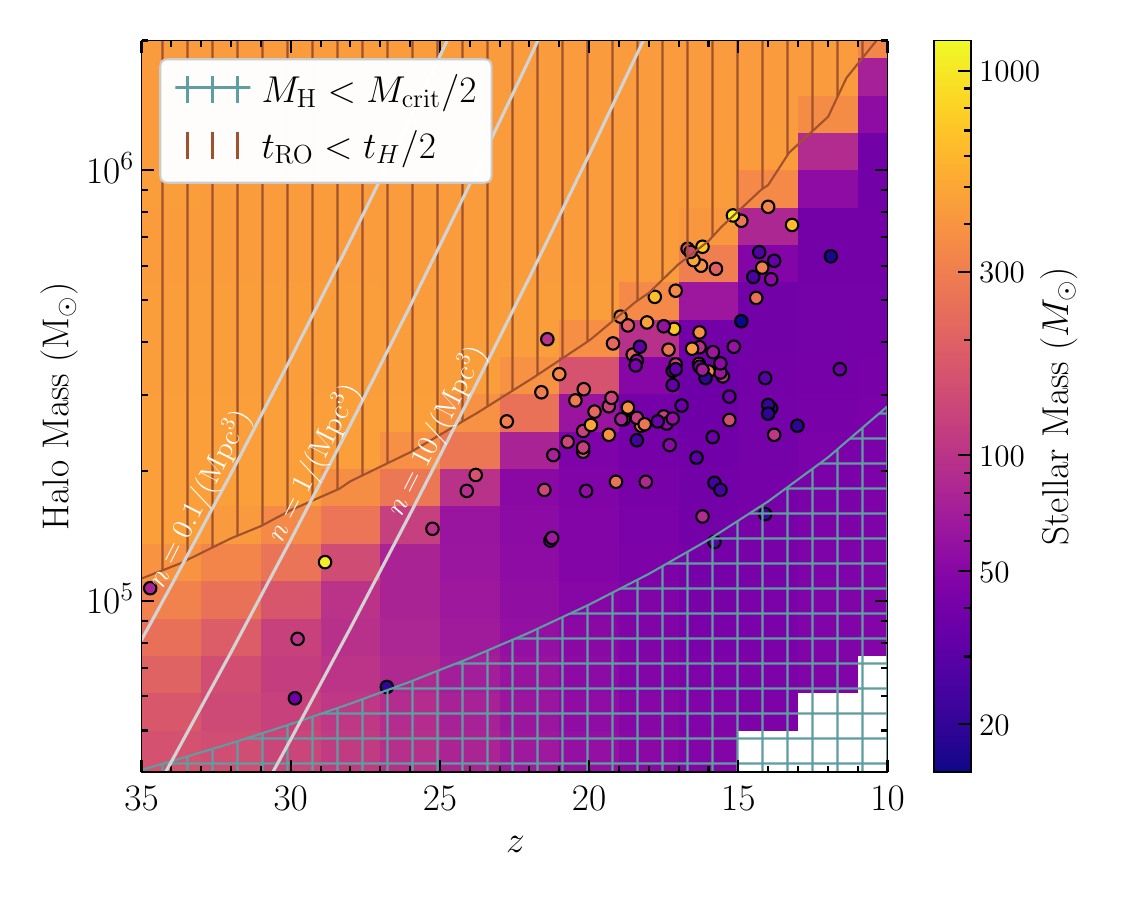}
    \caption{The stellar masses predicted by the combined model of this work (background color) assuming no fragmentation as in \cref{sec:singlestar}, compared with the simulations of \protect\cite{hirano_one_2014} (colored points). The transition from orange to purple is due to the delayed collapse and concomitant turn-on of $\hd$ cooling in the slowly cooling halos, as determined by \cref{eq:ro} (compare with \cref{fig:T_min}). In the bottom right, the virial temperature is so low that $\hto$ transitions cannot cool the halo and $t_{\rm RO}$ cannot be calculated using the prescription described in Section~\ref{sec:haloscale}. Also shown are the critical halo mass for Pop.~III star formation of \protect\cite{Kulkarni_2021} (teal, cross hatched), an upper mass limit for Pop.~III star formation based on the Rees-Ostriker condition (black, vertical hatched) and contours showing the integrated number of halos above a given mass (grey).}
    \label{fig:hiranocompare}
\end{figure}

The predicted stellar masses under the single-star fragmentation model and region of parameter space in which stars may form is shown in \cref{fig:hiranocompare}, along with the simulation sample of \cite{hirano_one_2014}. \cite{hirano_one_2014} used an azimuthal averaging during the accretion phase of their simulations, which prevented fragmentation of the disk. Also included are three contours of $n(>M_H)$, with the highest contour $0.1 \, \rm Mpc^{-3}$ approximately corresponding to the volume of the simulation box of \cite{hirano_one_2014}. Under equivalent assumptions (no disk fragmentation) the model agrees quite well with the simulation results. Especially, the transition from the $\hto$ mode to the $\hd$ mode in cooler halos is well reproduced. The parameters $\lambda$ and $\cal M$ are held fixed at their fiducial values (Table~\ref{tab:fid_pars}), while these values (and also $f$) vary even at fixed halo mass and redshift in the simulation results. The variation in these parameters explains some of the scatter of the simulation results compared to the model prediction. Although the model has several calibrated parameters (\cref{tab:calib_pars}), all take physically reasonable values which are not excessively fine tuned.

This concludes the validation of the model, since no simulations currently exist which probe a volume comparable to that of \cite{hirano_one_2014} with resolution adequate to resolve the fragmentation in the protostellar disk or with non-zero $J_{21}$. 

\section{Case Study: The Efficiency of Pop.~III Star Formation}
\label{sec:sfe}

We have argued that the close agreement in \cref{fig:hiranocompare} should be taken seriously despite the omission of key processes in the simulations of \cite{hirano_one_2014} because the SFE is relatively robust to these omissions. In this section, we demonstrate the utility of the model by studying the SFE in a wide range of cosmological environments, specifically including cases with non-zero Lyman-Werner intensity. Of the cosmological parameters in the model,  $J_{21}$ is the most recently incorporated (\cref{sec:haloscale}) and the least explored in the context of full halo-to-disk modeling; its effects on the star formation efficiency have not been simulated exhaustively across the range of halo masses and redshifts. The same procedure demonstrated in this section applies directly to any other parameter in \cref{tab:fid_pars}: the rotation parameter, the turbulent Mach number, or the fragmentation spectrum can each be varied across the cosmological parameter space in the same way. We focus on $J_{21}$ here because it cleanly illustrates how the framework connects a single change in the large-scale environment to quantitative shifts in the star formation outcome, without requiring new simulations at each point in parameter space.

We introduce two definitions of the SFE. The first is the cloud-scale SFE: the fraction of gravitationally unstable gas which is actually converted into stars (rather than being ejected by feedback). The cloud mass $M_c$ is defined as the mass at which $M/M_{\rm MBE}$ attains its maximum, and the SFE is then 
\begin{equation}
    \varepsilon_{{\rm SFE},c} = \frac{M_*}{M_c}.
\end{equation}

A second definition of the SFE (which is useful especially in numerical methods where individual star-forming clouds are unresolved) is the halo-scale SFE:
\begin{equation}
    \varepsilon_{{\rm SFE},\rm H} \equiv \frac{M_*}{\Omega _BM_{\rm H}},
\end{equation}
where $\Omega_b$ is the cosmic baryon fraction and $M_{\rm H}$ is the halo mass. These quantities are shown in \cref{tab:nt_sfe} for the four cases introduced in \cref{sec:haloscale} (\cref{fig:n-t_tracks}) and developed in the subsequent analysis. In two cases, $\varepsilon_{\rm SFE} \gtrsim 1$. The first is the cloud-scale SFE for the $\hd$ cooling example. This indicates only that accretion continues (albeit more slowly) beyond the peak of $M/M_{\rm MBE}$ (see \cref{fig:var_jlw_nt}). The second is the halo-scale SFE for the atomic cooling halo. This is due to the fact that this halo is in fact gas dominated due to the specific dark matter density profile we have adopted \cref{eq:dmsim}, while the definition of $\varepsilon_{SFE,H}$ assumes the cosmic baryon fraction. The cloud-scale SFE varies by a factor of few between the examples, and is of order unity. The halo scale SFE, in contrast, varies by more than two orders of magnitude. This large range is because the halo mass is only indirectly tied to the relevant cooling mechanism and thus the star formation outcomes. 

\begin{table*}
\centering
\caption{Cloud scale and halo scale SFE for the cases in \cref{fig:var_jlw_nt}, \cref{tab:nt_final}. See discussion in text of the cases with $\varepsilon \gtrsim 1$.}
\label{tab:cooling_sfe}
\renewcommand{\arraystretch}{1.5}
\begin{tabular}{lcccccc}
\hline
 & \textbf{$\hd$ cooling} &\textbf{\begin{tabular}[c]{@{}c@{}}$\hto$ cooling \\fiducial\end{tabular}} & \textbf{\begin{tabular}[c]{@{}c@{}}$\hto$ cooling \\ $\lambda=0.2$\end{tabular}} & \textbf{\begin{tabular}[c]{@{}c@{}}$\hto$ cooling\\  ${\cal M }=2$\end{tabular}} & \textbf{\begin{tabular}[c]{@{}c@{}}  H$_{2}$ cooling\\ $J_{21}=1$\end{tabular}} & \textbf{H cooling} \\ \hline

Cloud-scale SFE $\varepsilon_{\rm SFE,c}$ & 0.95 & 0.27 & 0.21 & 0.30 & .23 & .47 \\ 
Halo-scale SFE $\varepsilon_{{\rm SFE},\rm H}$ & 0.013 & 0.022 & 0.017 & 0.055 & $2.2 \times 10^{-3}$& 1.5 \\ 

\end{tabular}
\label{tab:nt_sfe}
\end{table*}

\begin{figure*}
     \centering
    \includegraphics[width=\linewidth,trim={0 7cm 0 0},clip]{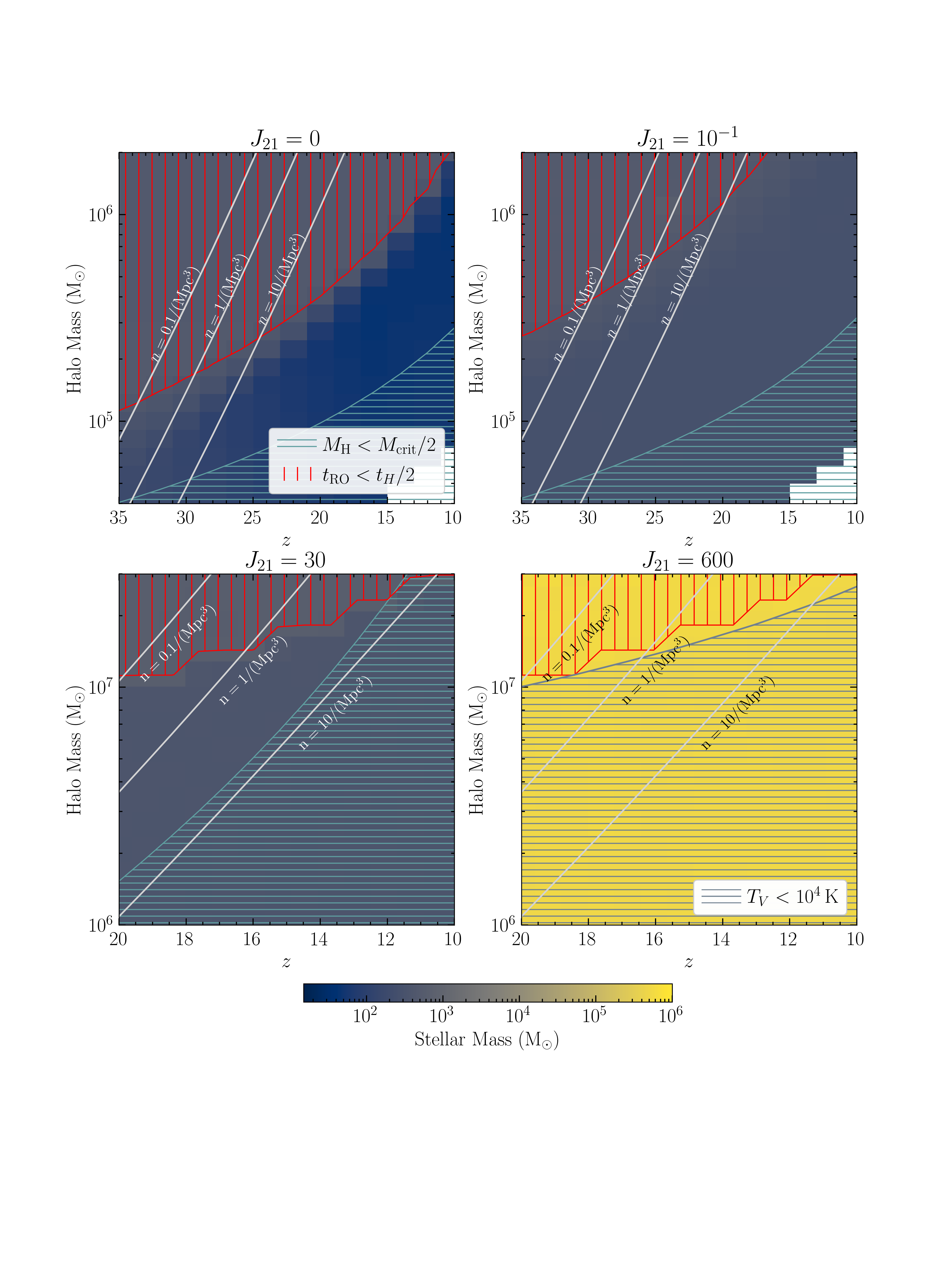}
    \caption{The predicted total stellar mass for different values of $J_{21}$  under the fiducial fragmentation model, as well as the bounds on the region where star formation is most likely to occur as in \cref{fig:hiranocompare}. Note the differing axes between the top (weaker radiation) and bottom (stronger radiation) cases, as well as the differing colorscale and dynamic range compared to \cref{fig:hiranocompare}. The top left panel differs from \cref{fig:hiranocompare} in that the fiducial fragmentation model is adopted here, in contrast to the single-star model in \cref{fig:hiranocompare} (although the difference is minor).}
    \label{fig:mstarcosmo}
\end{figure*}

 Now, we apply the model developed above to a broad range of halo masses and radiation intensities. Lyman-Werner radiation impedes the formation of $\hto$, which can delay or prevent gas collapse and alter the chemical-thermal evolution of the gas. In turn, this impacts the collapse and star formation dynamics as detailed in the preceding section. The specific impact of the radiation field on the chemical-thermal evolution of the collapsing gas depends on the details of the chemistry and radiative transfer as well as the halo growth history (since often $t_{\rm RO} \sim t_H$ for halos exposed to Lyman-Werner radiation). To study this dependence, we initialize the one-zone network as in \cref{sec:haloscale}, based on the radiation intensity, the cosmological freeze-out abundances, and the virial density and temperature. The thermal evolution then determines the infall rate via the fiducial cloud-scale parameters, from which we calculate the stellar mass using the fiducial disk model of \cref{sec:lowerlimit}. The total stellar mass for a range of halo masses and redshifts as well as the constraints from the halo mass function, Hubble time, and critical halo mass (for the appropriate $J_{21}$) are shown in \cref{fig:mstarcosmo} for a range of radiation intensities ($J_{21}$ =0, $J_{21}=0.1$, $J_{21}=30$ and $J_{21}=600$). As $J_{21}$ increases, the $\hd$ cooling mode is disabled in favor of $\hto$ and then $\rH$ cooling, and as the cooling time becomes longer the upper mass limit for Pop.~III star formation is pushed to higher halos masses (compare with \cref{fig:T_min}). For the stronger radiation intensities, we study larger halo masses and lower redshifts as low-mass halos become unable to cool at these intensities.

The critical halos mass of \cite{Kulkarni_2021} was calibrated only up to $J_{21}=30$. At $J_{21}=600$, the molecular channel is effectively shut off and cooling is impossible below the atomic limit temperature $T\sim 10^4 \, \rm K$. Thus, we use this as a lower limit for the mass of halos which form stars in the presence of strong dissociating radiation. Due to the efficiency of atomic cooling, the condition $t_{\rm RO}<t_H/2$ also aligns closely with the atomic limit temperature. The implication is that for strong radiation fields, Pop.~III star formation is restricted to those halos which are just reaching the atomic limit temperature. Between the four radiation intensities shown in \cref{fig:mstarcosmo}, the stellar mass varies from $\sim 50 \, \rm M_\odot$ to $\sim5\times 10^5 \, \rm M_\odot$. At the intermediate radiation intensities, the $\hd$ cooling mode which enable the formation of low-mass $\lesssim 100 \, \rm M_\odot$ clusters is shut down. In this regime, the stellar mass hardly depends on the halo properties. Meanwhile, at $J_{21}=600$ only $\rH$ cooling operates and the stellar mass is $\sim 10^5 \, \rm M_\odot$. 

\cite{nishijima2023lowmasspopiiistar} simulated a single cosmological initial condition subjected to a range of Lyman-Werner intensities. As the Lyman-Werner intensity increased, the collapse was delayed and the halo mass increased, effectively tracing a single contour in halo mass and redshift space. That work used a stiff equation of state to halt the collapse at a threshold density of $10^8 \, \rm cm^{-3}$, and did not follow the accretion phase of the evolution. \cite{nishijima2023lowmasspopiiistar} found a complicated dependence of the operant cooling mechanism (and implied stellar mass) on the radiation intensity, with the $\hd$ mode operating at $J_{21}=0$ and then strongly from $0.02 <J_{21}<0.1$ and more weakly from $0.1 <J_{21} < 10$. For $J_{21} =30$, atomic cooling was activated. However, the properties of the cloud for $J_{21}=0$ make this case an extreme outlier in the context of broader cosmological samples \cite[e.g.][]{hirano_one_2014,Hirano2015}, collapsing at redshift 25 with a virial mass of $\sim 10^6 \, \rm M_\odot$ and a total cloud mass of only $\sim 4 \, \rm M_\odot$. It is unclear at present how to interpret these results for more typical halos and clouds. 

Still, the general trend in \cref{fig:mstarcosmo} of diminished $\hd$ cooling in favor of $\hto$ and eventually $\rH$ at increasing radiation intensities is reasonable and broadly consistent with \cite{nishijima2023lowmasspopiiistar}. With higher $J_{21}$, the $\hto$ fraction is decreased at each density, which means that the evolution along the density-temperature trajectory $t_{\cc} \sim t_{\rm ff}$ is at higher temperature. Since $\hd$ fractionation can occur only near the $\hto$ temperature floor $\sim 200 \, \rm K$, this mode is shut off by modestly decreasing the $\hto$ cooling efficiency. The findings of \cite{nishijima2023lowmasspopiiistar} that intermediate $J_{21}$ enhance the $\hd$ channel can be understood if there is a window where the dissociating radiation slows down the collapse (one necessary ingredient for $\hd$ cooling) without ultimately decreasing too much the efficiency of the $\hto$ cooling around the critical density (the other necessary ingredient). In our simplified treatment, a Lyman-Werner field strong enough to delay the collapse ends up always decreasing the cooling efficiency at the critical density such that $\hd$ cannot efficiently form.

\begin{figure*}
     \centering
    \includegraphics[width=\linewidth,trim={0 7cm 0 0},clip]{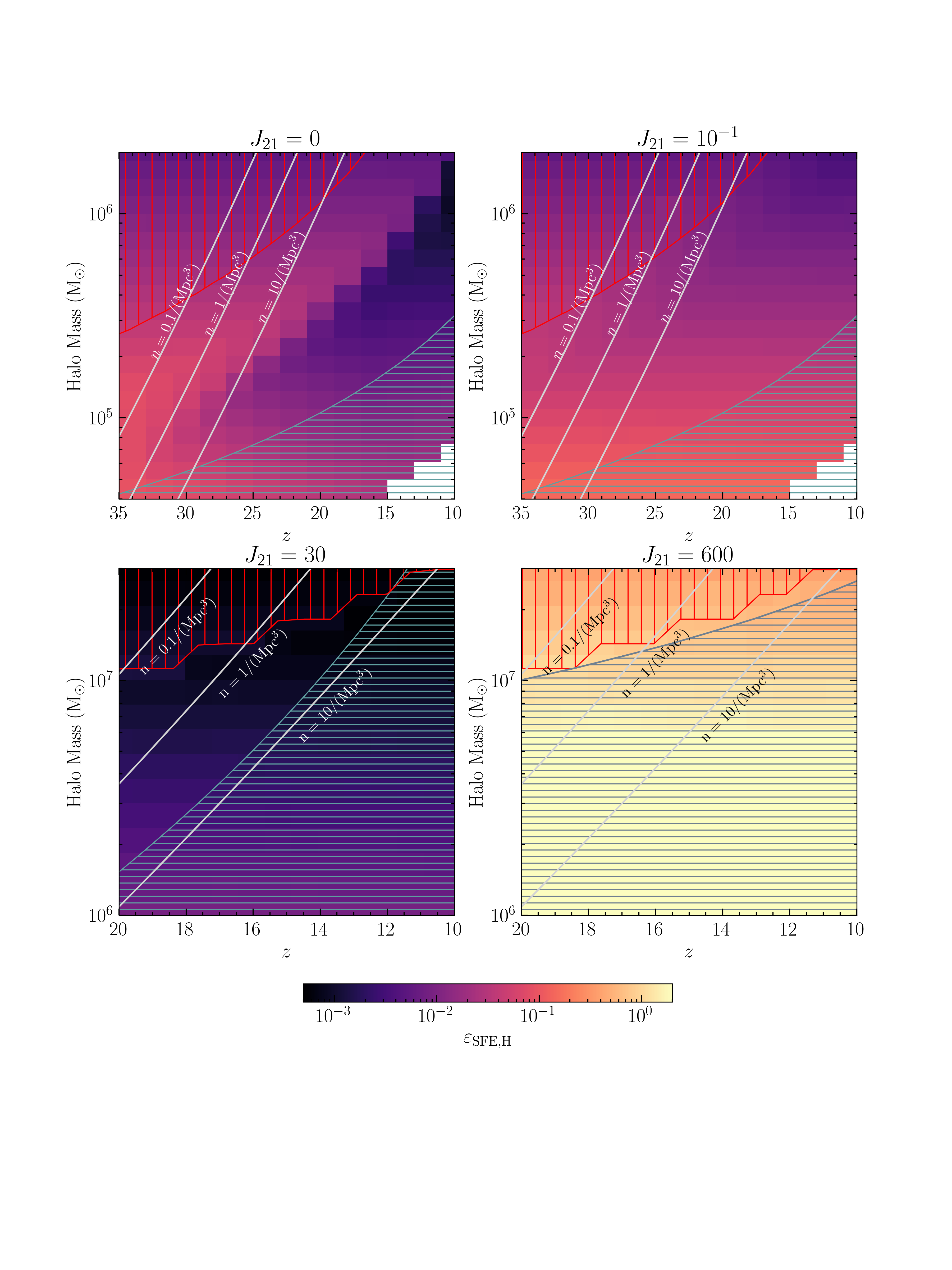}
    \caption{The halo-scale SFE for the same cases as \cref{fig:mstarcosmo}. This quantity varies by more than two orders of magnitude over the parameters studied here.}
    \label{fig:sfe_h}
\end{figure*}

Now, from these cluster masses it is straightforward to calculate the cloud- and halo-scale SFE.  \Cref{fig:sfe_h}, shows the halo scale SFE, $\varepsilon_{\rm SFE,c}$ for the same four radiation intensities. Even within the $J_{21}=0$ case, this quantity varies by about two orders of magnitude, from $\sim 10^{-3}$ in the highest mass $\hd$ cooled halos to $\sim 0.1$ in the lowest mass $\hto$ cooled halos. Across the full range of parameters, the variation is close to three orders of magnitude. The large variation is due to the relevant physical scale in the problem (the cloud mass dictated by the operant coolants) aliasing against the halo mass, which enters only indirectly.

\begin{figure*}
     \centering
    \includegraphics[width=\linewidth,trim={0 7cm 0 0},clip]{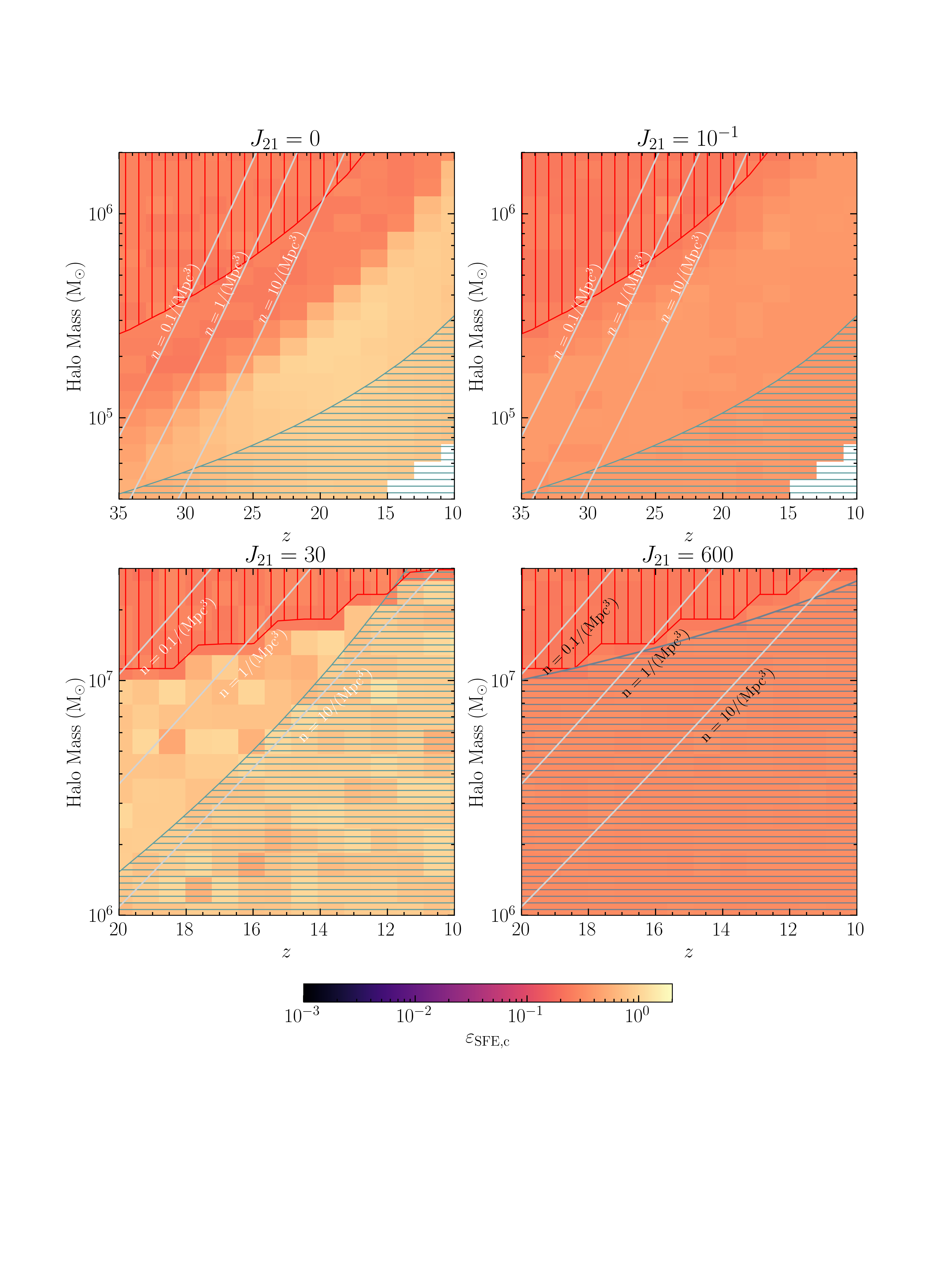}
    \caption{The cloud-scale SFE for the same cases as \cref{fig:mstarcosmo}. This quantity varies by only a factor of a few over the parameters studied here, compared to multiple decades for the halo-scale SFE (\cref{fig:sfe_h}).  The speckling in the bottom left is due to the flat shape of $M/M_{\rm MBE}$ near its peak for these cases, which leads to variation in the calculated cloud masses.}
    \label{fig:sfe_c}
\end{figure*}

Meanwhile, \cref{fig:sfe_c} shows the cloud-scale SFE for the same parameters. In contrast to the halo-scale SFE, this quantity varies by only a factor of few, in the range of $\sim 0.3\lesssim \varepsilon_{\rm SFE,c} \lesssim 1$. In this more physical definition of the SFE, Pop.~III star formation is highly efficient. That is, a relatively high fraction of the gravitationally unstable reservoir of gas is converted into stars. For Pop.~III star formation, the infall time ($\lesssim \rm 10^5 \, \rm yr$) is typically shorter than or comparable to the stellar lifetime ($\sim \rm Myr$), which leaves only direct stellar radiation as a feedback mode. In pristine gas, there is minimal opacity at wavelengths long-wards of Lyman-alpha. At the high accretion rates typical of Pop.~III star formation even the Kelvin-Helmholtz timescale can be a significant fraction of the infall timescale. Therefore, protostars spend much of the accretion phase with large radii and low surface temperatures, emitting negligible UV radiation. By this argument, the conclusion that Pop.~III star formation is efficient at the cloud scale is highly credible despite the uncertainties in the details of the accretion phase.

A corollary of these calculations is that the halo-scale efficiency of Pop.~III star formation is dictated by the efficiency with which limited coolants are able to convert the baryonic reservoir in a halo into a gravitationally unstable cloud.\footnote{This is in contrast to present day star formation, which is  regulated principally by feedback.} In other words, the efficiency of Pop.~III star formation in a given environment is mainly dictated by cloud-scale processes which are controlled indirectly by the halo-scale and cosmological environment.

\section{Summary and Conclusions}
\label{sec:discussion}
We have integrated \cite{gurian2024zero,liu2024,Gurian2025} into an end-to-end model of Pop.~III star formation. The combined model reproduces well the results of state-of-the art simulations at all relevant scales (halo, cloud, and disk). The analysis clearly illustrates the enormous dynamic range involved in Pop.~III star formation. Due to the top-heavy mass function produced by fragmentation in Pop.~III protostellar accretion disks, the local IMF is likely to be sensitive to the cosmological environment, which, together with the rate of fragmentation, determines an upper limit of the mass function. Fortunately, this otherwise intractable dynamic range can be handled by factorizing the problem into separate scales which can be modeled independently, with the influence of each larger scale on the smaller captured by a small number of parameters.

Further, we have shown that the efficiency of Pop.~III star formation is remarkably robust to the details of the fragmentation and feedback in the protostellar accretion disk \citep[at least, within the parameterizations of these processes adopted here and in ][]{liu2024}. This finding is consistent with simulations \citep{Prole2022, Sugimura2023}. This insensitivity is due to the importance of the cloud-scale infall rate in regulating the growth of protostars. While previous work has emphasized the importance of radiative feedback in shutting down accretion, we emphasize a more subtle interplay between the gas supply constraint and the ionizing feedback. In many cases, it is only as the cloud-scale infall rate drops precipitously at late times that photo-evaporation is able to balance and eventually overcome the rate at which gas is fed to the disk. In other words, although the proximal cause of the shutdown of accretion is the photo-evaporation of the disk, the root cause (especially in the low-mass $\hd$ cooled clouds) is more likely to be the depletion of the gravitationally unstable gas. Indeed, at the cloud scale, the SFE is typically tens of percent.

On the other hand, the halo-scale SFE varies by more than two orders of magnitude based on the large-scale environment of the particular star-forming cloud. The variation is due to the wide range in both cloud masses and halo masses. The lowest SFE $\varepsilon_{\rm SFE} \approx 10^{-3}$ occurs in the most massive halos cooled by $\hd$ (at negligible $J_{21}$) or the most massive halos cooled by $\hto$ (at stronger radiation intensity). Meanwhile, the highest SFE occurs in the low mass $\hto$ cooled halos and in the atomic cooling halos. The basic reason for the wide range in the halo-scale SFE is that the cloud mass is dictated primarily by the cooling mechanism, which depends only indirectly on the halo mass. However, the halo mass directly enters the (halo-scale) SFE. Because the cloud mass is determined primarily by gravity, chemistry, and optically thin radiative cooling (all of which can today be simulated at relatively high fidelity), these conclusions are likely robust to the small-scale uncertainties related to feedback, magnetic fields, and fragmentation. 

A related point is that it is critical for future high-resolution simulations of Pop.~III star formation to consider the fidelity of the cloud-scale initial conditions. It is common practice to adopt an isothermal Bonnor-Ebert sphere with the density everywhere enhanced by some order unity factor (in order to promote gravitational collapse) as the initial condition for such simulations. This work illustrates that the ratio of the local density to the critical Bonnor-Ebert sphere is not in fact a free parameter, but is instead dictated by the early chemical-thermal evolution. Interestingly, the 1D simulation of \cite{Omukai2010} (which adopted such a Bonnor-Ebert initial condition) adjusted by the first snapshot to a density profile very similar to the one predicted by the model of \cref{sec:cloud_scale} \citep{Gurian2025}. However, the initial conditions leave an imprint in the velocity structure of the cloud at large masses/radii, which may play a critical role in setting the final mass of the stellar cluster. 

In future work, this model can be extended or applied in several ways. Perhaps most immediately, we have so far neglected the relative velocity between baryons and dark matter, which has been shown to profoundly impact the Pop.~III star formation processes \citep[e.g.][]{Stacy2011,Latif2014,Hirano_2017, Schauer_2019,nakazato_h_2022, Lake_2023, Lake_2025, Hirano2025}. Some work suggests that the separation of scales discussed in this work is not as clean in regions of high streaming velocity, where the gas clouds can undergo turbulent fragmentation rather than forming a rotationally supported disk. Still, the basic approach of this work can be applied as long as the effects of the streaming velocity on each individual scale can be modeled. This application is especially interesting as a point of comparison to present day star formation, with the extremely high gas temperatures and infall rates typical of Pop.~III star formation but the strong turbulence of present day star formation. 

Similarly, JWST observations suggest that the SFE is likely higher at high redshift than in the local universe, perhaps much higher \citep{harikane2023purespectroscopicconstraintsuv,robertson2024earliestgalaxiesjadesorigins,Chemerynska_2026}. In this context, the Pop.~III case studied here is a limiting case where the SFE is again often low (at the halo scale), but due to the inefficiency of cooling rather than the efficiency of feedback. The question of the transition of star formation out of the Pop.~III mode is old and deep \citep{Omukai_2005,Chon_2021,Sharda22,Chon_2024}. It would be interesting to apply some of the machinery developed here to that problem. 

Further, the essential input to the cloud and disk scale models is a density-temperature relationship, which we here determined using one-zone models. For a large delay factor $f$, the cooling timescale becomes comparable to or longer than the Hubble time. In that case, it is important in principle to follow the growth of the halo across cosmic time. That is, our assumption of a single, well-defined virial density and temperature at which the one-zone model can be initialized breaks down. Dynamical heating caused by mergers and rapid halo assembly can affect the gas cloud properties and the degree of turbulence \citep{Yoshida03}.
These effects can be more realistically accounted for by using semi-analytic merger trees to track the chemical-thermal history of the gas before the onset of runaway collapse, similar to \cite{Li_2021}. In this context, the ``critical halo mass'' for Pop.~III star formation can be tracked on a halo-by-halo basis to the onset of the Rees-Ostriker instability. 

The disk-scale fragmentation model adopted here is phenomenological: the power-law mass function, the fixed lower cutoff, and the ansatz for the growth of the upper mass limit are parameterizations of processes that no existing simulation resolves self-consistently through the termination of accretion. A first-principles theory of the fragmentation cascade in Pop III protostellar disks, connecting the opacity-limited Jeans mass to the final mass function without free parameters, remains an open problem. The insensitivity of the total stellar mass to the fragmentation parameters (\cref{sec:lowerlimit,fig:alpha_compare}) suggests that such a theory is not required for predicting the star formation efficiency, but it is essential for predicting the Pop III IMF itself. Progress on this front, whether through higher-resolution simulations that converge on the low-mass end or through analytic models of angular momentum transport and competitive accretion in self-gravitating disks, would directly slot into the framework developed here by replacing the parameterized disk model with a predictive one.

Finally, simulations are emerging which include Pop.~III stars (in some cases tracking individual stars) but only marginally resolve the star-forming clouds \citep[e.g.][]{Brauer_2025,Brauer_2025a,Zier_2025}. These simulations currently adopt a universal, ad-hoc IMF for Pop.~III stars. Instead, the cloud and disk scale models described here could be adopted as subgrid prescriptions in such simulations, revealing the back-reaction of sub-parsec-scale Pop.~III physics on cosmological observables.

\section*{Acknowledgments}
We thank Kazuyuki Sugimura for sharing data products from \cite{Sugimura2023}. BL acknowledges the funding of the Deutsche Forschungsgemeinschaft (DFG, German Research Foundation) under Germany's Excellence Strategy EXC 2181/1 - 390900948 (the Heidelberg STRUCTURES Excellence Cluster). DJ was supported by NSF grant AST-2407298 at PSU, as well as by a KIAS Individual Grant, PG088301. TH acknowledges the financial support from the JSPS (19KK0353 and 22H00149) and the Kyoto University Foundation. SH acknowledges the financial support from the JSPS (21K13960, 21H01123, and 23K20864). NY thanks financial support from JSPS KAKENHI International Leading Research 23K20035 and Specially Promoted Research 24H00004. Research at the Perimeter Institute is supported in part by the Government of Canada through the Department of Innovation, Science and Economic Development Canada and by the Province of Ontario through the Ministry of Colleges and Universities.

\section*{Data Availability}
The code and data underlying this paper will be shared on reasonable request to JG at \url{jgurian@perimeterinstitute.ca}. 

\bibliographystyle{mnras}
\bibliography{popiii}
\appendix
\section{Derivation of Accretion Rate}
\label{app:mdot}
Consider the mass function 
\begin{equation}
    \frac{dN}{dm} = c(t) m^\alpha \mbox{\ ,}
    \label{eq:mf}
\end{equation}
where $c(t)$ fixes the total mass in stars
\begin{equation}
    M_*=\int_{m_1^+}^{m_2^-}dm\,\frac{dN}{dm} m \mbox{\ .}
    \label{eq:mftot}
\end{equation}
Here, we are avoiding the undefined $dN/dm$ at the boundary by taking the limits towards $m_1$ and $m_2$ from above and below, respectively.

Now, the continuity equation in mass space is

\begin{equation}
    \frac{\partial }{\partial t}\frac{dN}{dm}+\frac{\partial  }{\partial m}\left(\dot m\frac{dN}{dm}\right)=j,
\end{equation}
with $j$ a source term. In principle this equation could be integrated along characteristics to self-consistently evolve the mass function. However, lacking clear insight on the form which $j$ ought to take, we assume that protostars are produced only at $m_1$:
\begin{equation}
    j = \dot N \delta(m-m_1)
\end{equation}

Differentiating \cref{eq:mftot} gives the total accretion rate
\begin{align}
\begin{split}
        \dot M_* &=\int_{m_1^+}^{m_2^-}dm\,\frac{\partial}{\partial t}\left(\frac{dN}{dm} \right)m \\ &+ \left[ m_2\frac{dN}{dm}\bigg|_{m_2^-}\frac{\partial m_2}{\partial t}  - m_1\frac{dN}{dm}\bigg|_{m_1^+}\frac{\partial m_1}{\partial t}\right]
        \end{split}
\end{align}

Inserting the continuity equation (and noting that the source term is identically zero in the domain of integration)

\begin{equation}
\begin{split}
\dot M_* &= \int_{m_1^+}^{m_2^-} dm \, m\left[- \frac{\partial}{\partial m} \left(\dot m\frac{dN}{dm}\right)\right] \\
&+ \left[m_2\frac{dN}{dm}\bigg|_{m_2^-}\frac{\partial m_2}{\partial t}  - m_1\frac{dN}{dm}\bigg|_{m_1^+}\frac{\partial m_1}{\partial t}\right].
\end{split}
\end{equation}

This expression can be integrated by parts
\begin{equation}
\begin{split}
    \dot M_* &= \int_{m_1^+}^{m_2^-} dm \,\dot m \frac{dN}{dm }   - \left[m \dot m \frac{dN}{dm}\right]_{m_1^+}^{m_2^-}\\ &+ \left[m_2\frac{dN}{dm}\bigg|_{m_2^-}\frac{\partial m_2}{\partial t}  - m_1\frac{dN}{dm}\bigg|_{m_1^+}\frac{\partial m_1}{\partial t}\right]. 
    \end{split}
\end{equation}

Note that the first bracketed term after the integral depends on the accretion rate at the boundary, while the second term depends on the time derivative of the boundary. We take 
\begin{align}
    \frac{\partial m_2}{\partial t} &= \dot m_2\\
    \frac{\partial m_1}{\partial t} &= 0,\\
\end{align}
which is to say that the upper boundary evolves according to the accretion rate at the upper limit, while the lower boundary is fixed. Then,
\begin{equation}
\begin{split}
    \dot M_* &= \int_{m_1^+}^{m_2^-} dm \,\dot m \frac{dN}{dm }   +m_1 \dot m_1 \frac{dN}{dm}\bigg|_{m_1^+} ,
    \end{split}
\end{equation}

from which we identify
\begin{equation}
     \dot m_1 \frac{dN}{dm}\bigg|_{m_1^+} = \dot N.
\end{equation}

We turn now to the form of $\dot m$. The continuity equation can be integrated after inserting \cref{eq:mf}, yielding:  
\begin{equation}
    \dot m = m^{-\alpha}K(t) - m \frac{\dot c(t)}{(1+\alpha) c(t)}.
    \label{eq:mdotofm}
\end{equation}

This relation comprises two terms: one proportional to $m$ and the other to $m^{-\alpha}$, as expected for self-similar growth. Note that for $\dot c =0$ we recover the standard competitive accretion result $\dot m \propto m^{-\alpha}$. The constraint on the total accretion rate then yields
\begin{equation}
  K(t) = \frac{1}{m_2 c(t)}\left[\dot M_* + \frac{\dot c}{2+\alpha} \left(m_1^{2+\alpha} + \frac{m_2^{2+\alpha}}{1+\alpha}\right)\right]
\end{equation}

Now, the constraint on total mass fixes the normalization as
\begin{equation}
    c(t)= \frac{M_*(\alpha+2)}{m_2^{\alpha+2}\left[1-\left(\frac{m_1}{m_2}\right)^{\alpha+2}\right]},
\end{equation}
so that also
\begin{equation}
    \dot c(t)= \left(\frac{\dot M_*}{M_*}-\frac{m_2^{1+\alpha}c(t)}{M_*}\frac{\partial m_2}{\partial t}\right)c(t).
\end{equation}

Physically, in this model the accretion rate is a sum of an exponential self-similar dilation term  $\dot m \propto m$ and a zero mass flux ``competitive'' term $\dot m \propto m^{-\alpha}$. The former can be realized if each protostar is fed by a disk whose mass is proportional to the proto-stellar mass on a common viscous timescale, while the latter could be realized by Bondi-Hoyle accretion. However, in those scenarios the overall normalization would be directly fixed by the accretion mechanism while in this case the normalization is written in terms of the global accretion constraint.

At this point, $\dot m_2=\partial m_2/\partial t$ remains free: evaluating \cref{eq:mdotofm} at $m_2$ yields a tautology. However, it is possible to express $\dot m_2$ in terms of $\dot N$, the rate at which protostars are produced at $m_1$. By writing 
\begin{equation}
    \dot m_2 m^{\alpha}c(t) - \dot m_1 m^\alpha c(t) = \dot m_2 m^\alpha c(t) - \dot N,
\end{equation}
expanding the accretion law \cref{eq:mdotofm} in the left hand side, and substituting \cref{eq:cdot}, one can show that 
\begin{equation}
    \dot m_2 \frac{dN}{dm}\bigg|_{m_2^-} = \frac{\dot M_* - \bar m \dot N}{m_2 - \bar m},
    \label{eq:m2ofndot}
\end{equation}
where 
\begin{equation}
    \bar m = \frac{\int_{m_1}^{m_2}dm \, m\frac{dN}{dm} }{\int_{m_1}^{m_2}dm \, \frac{dN}{dm}}.
\end{equation}

\bsp	% typesetting comment
\label{lastpage}
\end{document}